\documentclass[aps,prl,superscriptaddress,showpacs,byrevtex]{revtex4}

\usepackage{graphicx} 
\usepackage{dcolumn}  

\newcommand{\bbbar}{\ensuremath{B\bar{B}}}
\newcommand{\qqbar}{\ensuremath{q\bar{q}}}

\newcommand{\bckppos} {\ensuremath{B^+\to K^+\pi^+\pi^-}}
\newcommand{\bckppss} {\ensuremath{B^+\to K^-\pi^+\pi^+}}
\newcommand{\bckkpos} {\ensuremath{B^+\to K^+K^-\pi^+}}
\newcommand{\bckkpss} {\ensuremath{B^+\to K^+K^+\pi^-}}
\newcommand{\bckkk}   {\ensuremath{B^+\to K^+K^+K^-}}
\newcommand{\bcksksp} {\ensuremath{B^+\to K^0_SK^0_S\pi^+}}
\newcommand{\bcksksk} {\ensuremath{B^+\to K^+K^0_SK^0_S}}
\newcommand{\bckknkn} {\ensuremath{B^+\to K^+K^0\bar{K}^0}}

\newcommand{\bnkspp}  {\ensuremath{B^0\to K^0_S\pi^+\pi^-}}
\newcommand{\bnknkp}  {\ensuremath{B^0\to K^0K^+\pi^-}}

\newcommand{\bnknkk}  {\ensuremath{B^0\to K^0K^+K^-}}
\newcommand{\bnkskk}  {\ensuremath{B^0\to K^0_SK^+K^-}}
\newcommand{\bnksksks}{\ensuremath{B^0\to K^0_SK^0_SK^0_S}}
\newcommand{\bnknknkn}{\ensuremath{B^0\to K^0K^0\bar{K}^0}}

\newcommand{\bcnkcnpp}{\ensuremath{B^{+(0)}\to K^{+(0)}\pi^+\pi^-}}
\newcommand{\bcnkcnkp}{\ensuremath{B^{+(0)}\to K^{+(0)}K^-\pi^+}}
\newcommand{\bcnkcnkk}{\ensuremath{B^{+(0)}\to K^{+(0)}K^+K^-}}

\newcommand{\kppos} {\ensuremath{K^+\pi^+\pi^-}}
\newcommand{\kppss} {\ensuremath{K^-\pi^+\pi^+}}
\newcommand{\kkpos} {\ensuremath{K^+K^-\pi^+}}
\newcommand{\kkpss} {\ensuremath{K^+K^+\pi^-}}
\newcommand{\kkk}   {\ensuremath{K^+K^+K^-}}
\newcommand{\ksksp} {\ensuremath{K^0_SK^0_S\pi^+}}
\newcommand{\ksksk} {\ensuremath{K^+K^0_SK^0_S}}
\newcommand{\klklk} {\ensuremath{K^+K^0_LK^0_L}}
\newcommand{\ksklk} {\ensuremath{K^+K^0_SK^0_L}}
\newcommand{\knknk} {\ensuremath{K^+K^0\bar{K}^0}}

\newcommand{\knpp}  {\ensuremath{K^0\pi^+\pi^-}}

\newcommand{\knkp}  {\ensuremath{K^0K^+\pi^-}}
\newcommand{\kskp}  {\ensuremath{K^0_SK^+\pi^-}}
\newcommand{\knkk}  {\ensuremath{K^0K^+K^-}}
\newcommand{\kskk}  {\ensuremath{K^0_SK^+K^-}}
\newcommand{\ksksks}{\ensuremath{K^0_SK^0_SK^0_S}}

\newcommand{\kcnpp}{\ensuremath{K^{+(0)}\pi^+\pi^-}}
\newcommand{\kcnkp}{\ensuremath{K^{+(0)}K^-\pi^+}}
\newcommand{\kcnkk}{\ensuremath{K^{+(0)}K^+K^-}}

\newcommand{\pipi}{\ensuremath{\pi^+\pi^-}}
\newcommand{\kpkm}{\ensuremath{K^+K^-}}

\newcommand{\ksks}{\ensuremath{K^0_SK^0_S}}
\newcommand{\klkl}{\ensuremath{K^0_LK^0_L}}
\newcommand{\kskl}{\ensuremath{K^0_SK^0_L}}
\newcommand{\knkn}{\ensuremath{K^0\bar{K}^0}}

\newcommand{\de}{\ensuremath{\Delta E}}
\newcommand{\mb}{\ensuremath{M_{\rm bc}}}

\newcommand{\bfkppos}  {\ensuremath{53.6\pm3.1\pm5.1}}
\newcommand{\bfknpp}   {\ensuremath{45.4\pm5.2\pm5.9}}
\newcommand{\bfkkk}    {\ensuremath{32.8\pm1.8\pm2.8}}
\newcommand{\bfknkk}   {\ensuremath{28.3\pm3.3\pm4.0}}
\newcommand{\bfksksk}  {\ensuremath{13.4\pm1.9\pm1.5}}
\newcommand{\bfksksks} {\ensuremath{4.2^{+1.6}_{-1.3}\pm0.8}}
\newcommand{\bfkkpos}  {\ensuremath{9.3\pm2.3\pm1.1}}
\newcommand{\ulkkpos}  {\ensuremath{<13}}
\newcommand{\ulknkp}   {\ensuremath{<18}}
\newcommand{\ulksksp}  {\ensuremath{<3.2}}
\newcommand{\ulkppss}  {\ensuremath{<4.5}}
\newcommand{\ulkkpss}  {\ensuremath{<2.4}}

\begin{document}


\title{      Study of $B$ meson decays to three-body charmless
                          hadronic final states}

\affiliation{Aomori University, Aomori}
\affiliation{Budker Institute of Nuclear Physics, Novosibirsk}
\affiliation{Chiba University, Chiba}
\affiliation{Chuo University, Tokyo}
\affiliation{University of Cincinnati, Cincinnati, Ohio 45221}
\affiliation{University of Frankfurt, Frankfurt}
\affiliation{Gyeongsang National University, Chinju}
\affiliation{University of Hawaii, Honolulu, Hawaii 96822}
\affiliation{High Energy Accelerator Research Organization (KEK), Tsukuba}
\affiliation{Hiroshima Institute of Technology, Hiroshima}
\affiliation{Institute of High Energy Physics, Chinese Academy of Sciences, Beijing}
\affiliation{Institute of High Energy Physics, Vienna}
\affiliation{Institute for Theoretical and Experimental Physics, Moscow}
\affiliation{J. Stefan Institute, Ljubljana}
\affiliation{Kanagawa University, Yokohama}
\affiliation{Korea University, Seoul}
\affiliation{Kyoto University, Kyoto}
\affiliation{Kyungpook National University, Taegu}
\affiliation{Institut de Physique des Hautes \'Energies, Universit\'e de Lausanne, Lausanne}
\affiliation{University of Ljubljana, Ljubljana}
\affiliation{University of Maribor, Maribor}
\affiliation{University of Melbourne, Victoria}
\affiliation{Nagoya University, Nagoya}
\affiliation{Nara Women's University, Nara}
\affiliation{National Kaohsiung Normal University, Kaohsiung}
\affiliation{National Lien-Ho Institute of Technology, Miao Li}
\affiliation{Department of Physics, National Taiwan University, Taipei}
\affiliation{H. Niewodniczanski Institute of Nuclear Physics, Krakow}
\affiliation{Nihon Dental College, Niigata}
\affiliation{Niigata University, Niigata}
\affiliation{Osaka City University, Osaka}
\affiliation{Osaka University, Osaka}
\affiliation{Panjab University, Chandigarh}
\affiliation{Peking University, Beijing}
\affiliation{Princeton University, Princeton, New Jersey 08545}
\affiliation{RIKEN BNL Research Center, Upton, New York 11973}
\affiliation{Saga University, Saga}
\affiliation{University of Science and Technology of China, Hefei}
\affiliation{Seoul National University, Seoul}
\affiliation{Sungkyunkwan University, Suwon}
\affiliation{University of Sydney, Sydney NSW}
\affiliation{Tata Institute of Fundamental Research, Bombay}
\affiliation{Toho University, Funabashi}
\affiliation{Tohoku Gakuin University, Tagajo}
\affiliation{Tohoku University, Sendai}
\affiliation{Department of Physics, University of Tokyo, Tokyo}
\affiliation{Tokyo Institute of Technology, Tokyo}
\affiliation{Tokyo Metropolitan University, Tokyo}
\affiliation{Tokyo University of Agriculture and Technology, Tokyo}
\affiliation{Toyama National College of Maritime Technology, Toyama}
\affiliation{University of Tsukuba, Tsukuba}
\affiliation{Utkal University, Bhubaneswer}
\affiliation{Virginia Polytechnic Institute and State University, Blacksburg, Virginia 24061}
\affiliation{Yokkaichi University, Yokkaichi}
\affiliation{Yonsei University, Seoul}

\author{A.~Garmash}\affiliation{Budker Institute of Nuclear Physics, Novosibirsk}\affiliation{High Energy Accelerator Research Organization (KEK), Tsukuba} 
\author{K.~Abe}\affiliation{High Energy Accelerator Research Organization (KEK), Tsukuba} 
\author{K.~Abe}\affiliation{Tohoku Gakuin University, Tagajo} 
\author{T.~Abe}\affiliation{High Energy Accelerator Research Organization (KEK), Tsukuba} 
\author{I.~Adachi}\affiliation{High Energy Accelerator Research Organization (KEK), Tsukuba} 
\author{H.~Aihara}\affiliation{Department of Physics, University of Tokyo, Tokyo} 
\author{M.~Akatsu}\affiliation{Nagoya University, Nagoya} 
\author{Y.~Asano}\affiliation{University of Tsukuba, Tsukuba} 
\author{T.~Aushev}\affiliation{Institute for Theoretical and Experimental Physics, Moscow} 
\author{A.~M.~Bakich}\affiliation{University of Sydney, Sydney NSW} 
% ...
\collaboration{The Belle Collaboration}

\begin{abstract}
  We report results of a study of charmless $B$ meson decays to three-body
$K\pi\pi$, $KK\pi$ and $KKK$ final states. Measurements of branching
fractions for $B$ decays to $\kcnpp$, $\kkk$, $\knkk$, $\ksksk$ and $\ksksks$
final states are presented. The decays $\bnknkk$, $\bcksksk$ and $\bnksksks$
are observed for the first time. We also report evidence for $\bckkpos$
decay. For the three-body final states $\knkp$, $\ksksp$, $\kkpss$ and
$\kppss$ 90\% confidence level upper limits are reported. Finally, we discuss
the possibility of using the three-body $\bnkskk$ decay for CP violation
studies. The results are obtained with a 78\,fb$^{-1}$ data sample collected
at the $\Upsilon(4S)$ resonance by the Belle detector operating at the KEKB
asymmetric energy $e^+e^-$ collider.
\end{abstract}
\pacs{13.25.Hw, 14.40.Nd}

\maketitle

\section{Introduction}

Studies of three-body charmless hadronic final states are a natural extension
of studies of two-body final states. Some of the final states considered so
far as two-body (for example $\rho \pi$, $K^*\pi$, etc.) are, in fact,
quasi-two-body since they produce a wide resonance state that immediately
decays, in
the simplest case, into two particles producing a three-body final state.
Multiple resonances occurring nearby in phase space will interfere and a full
amplitude analysis is required to extract correct branching fractions for
the intermediate quasi-two-body states. $B$ meson decays to three-body
charmless hadronic final states may also provide new possibilities for CP
violation searches. For example, a new method to extract the weak angle
$\phi_3$ from isospin analysis and measurement of time dependent asymmetry in
the decay
$\bnkspp$ has been recently suggested in Ref.~\cite{b2hhhcp}. Charmless decays
of $B$ mesons are also important in current searches for physics beyond the
Standard Model (SM). Among three-body charmless final states, the $\bckkpss$
and $\bckppss$ decays, which proceed via $b\to ss\bar d$ and $b\to dd\bar s$
transitions, respectively, provide a good opportunity to search for new
physics. The SM prediction for the branching fraction of the decay $\bckkpss$
is of order $10^{-11}$ and even smaller for the $\kppss$ final
state~\cite{huiti-1}. However, there are extensions of the SM where these
branching fractions can be enhanced up to $10^{-7}$~\cite{huiti-1,huiti-2}.
Upper limits on branching fractions for these final states can be used to
constrain parameters in some extensions of the SM~\cite{fajfer}.

$B$ meson decays to $\kcnpp$ and $\kkk$ final states have already been observed
by the Belle, BaBar and CLEO experiments~\cite{garmash,aubert,eckhart}. For
the other three-body charmless hadronic final states considered in this paper
only upper limits have been reported so far~\cite{PDG}. In this paper, we
present updated results on a study of $B$ meson decays to three-body $K\pi\pi$,
$KK\pi$ and $KKK$ final states. We also describe an isospin analysis of the
decays of $B$ mesons to three-kaon final states and discuss the use of the
$\bnkskk$ three-body decay for CP violation measurements.

The analysis is based on a 78\,fb$^{-1}$ data sample, which contains 85.0
million $B\bar{B}$ pairs, collected  with the Belle detector  operating at the
KEKB asymmetric-energy $e^+e^-$ collider~\cite{KEKB} with a center-of-mass
(c.m.) energy at the $\Upsilon(4S)$ resonance. The beam energies are 3.5 GeV
for positrons and 8.0 GeV for electrons. For the study of the
$e^+e^-\to q\bar{q}$ continuum background we use 8.3\,fb$^{-1}$ of data taken
60~MeV below the $\Upsilon(4S)$ resonance. The results presented here
include the previous data and supersede the results on three-body charmless
hadronic final states reported in Ref.~\cite{garmash}.

\section{The Belle detector}

  The Belle detector~\cite{Belle} is a large-solid-angle magnetic spectrometer
based on a 1.5~T superconducting solenoid magnet. Charged particle tracking is
provided by a three-layer silicon vertex detector and a 50-layer central
drift chamber (CDC) that surround the interaction point. The charged particle
acceptance covers laboratory polar angle between $\theta=17^{\circ}$ and
$150^{\circ}$, corresponding to about 92\% of the full solid angle in the
c.m.\ frame. The momentum resolution is determined from cosmic
rays and $e^+ e^-\to\mu^+\mu^-$ events to be 
$\sigma_{p_t}/p_t = (0.30 \oplus 0.19 p_t)\%$, where $p_t$ is the transverse
momentum in GeV/$c$.

  Charged hadron identification is provided by $dE/dx$ measurements in the CDC,
an array of 1188 aerogel \v{C}erenkov counters (ACC), and a barrel-like array
of 128 time-of-flight scintillation counters (TOF); information from the three
subdetectors is combined to form a single likelihood ratio, which is then used
in kaon and pion selection. At large momenta
(\mbox{$>2.5$}~GeV/$c$) only the ACC and CDC are used to separate charged
pions and kaons since here the TOF provides no additional discrimination.
Electromagnetic showering particles are detected in an array of 8736 CsI(Tl)
crystals that covers the same solid angle as the charged particle tracking
system. The energy resolution for electromagnetic showers is
$\sigma_E/E = (1.3 \oplus 0.07/E \oplus 0.8/E^{1/4})\%$, where $E$ is in GeV.
Electron identification in Belle is based on a combination of $dE/dx$
measurements in the CDC, the response of the ACC, and the position, shape and
total energy deposition (i.e., $E/p$) of the shower detected in the
calorimeter. The electron identification efficiency is greater than 92\% for
tracks with $p_{\rm lab}>1.0$~GeV/$c$ and the hadron misidentification
probability is below 0.3\%.
The magnetic field is returned via an iron yoke that is instrumented to detect
muons and $K^0_L$ mesons. We use a GEANT-based Monte Carlo (MC) simulation to
model the response of the detector and determine acceptance~\cite{GEANT}.


\section{Event selection}

Charged tracks are selected with a set of track quality requirements based on
the average hit residual and on the distances of closest approach to the
interaction point (IP). We also require that the track momenta transverse to
the beam be
greater than 0.1~GeV/$c$ to reduce the low momentum combinatorial background.
For charged kaon identification, we impose a requirement on the particle
identification variable, which has 86\% efficiency and a 7\% fake rate from
misidentified pions. Charged tracks that are positively identified as electrons
or protons are excluded. Since the muon identification efficiency and fake rate
vary significantly with the track momentum, we do not veto muons to avoid
additional systematic errors.

Neutral kaons are reconstructed via the decay $K^0(\bar{K}^0)\to\pi^+\pi^-$.
The invariant mass of the two pions is required to be within 12~MeV/$c^2$ of
the nominal $K^0$ mass. The displacement of the $\pi^+\pi^-$ vertex from the
IP in the $r$-$\phi$ plane is required to be greater than 0.1~cm and less than
20~cm. The direction of the combined pion pair momentum in the $r$-$\phi$ plane
must be within 0.2 rad of the direction from the IP to the displaced vertex.

We reconstruct $B$ mesons in three-body $K\pi\pi$, $KK\pi$ and $KKK$ final
states, where $K$ stands for a charged or neutral kaon, and $\pi$ for a charged
pion. The inclusion of the charge conjugate mode is implied throughout this
report. The candidate events are identified by their c.m.\ energy difference,
$\de=(\sum_iE_i)-E_{\rm beam}$, and the beam constrained mass,
$\mb=\sqrt{E^2_{\rm beam}-(\sum_i\vec{p}_i)^2}$, where
$E_{\rm beam}=\sqrt{s}/2$ is the beam energy in the c.m.\ frame, and
$\vec{p}_i$ and $E_i$ are the c.m.\ three-momenta and energies of the candidate
$B$ meson decay products. We select events with $\mb>5.20$~GeV/$c^2$ and
$-0.30<\de<0.50$~GeV. For subsequent analysis, we also define a $\mb$
{\it signal} region of $|\mb-M_B|<9$~MeV/$c^2$.

To determine the signal yield, we use events with $\mb$ in the signal region
and fit the $\de$ distribution with the sum of a signal and a background
function. The $\de$ signal is parameterized by the sum of two Gaussian
functions with the same mean. The widths and the relative fractions of the two
Gaussians are determined from the MC simulation. We find that the signal MC
events have a 10\% narrower $\de$ width than seen in the data. To correct for
this effect, we introduce a scale factor that is determined from the comparison
of the $\de$ widths for $B^+\to\bar{D}^0\pi^+$ events in MC and experimental
data. The $\de$ shape of the $\bbbar$ produced background is determined from MC
simulation, as described below. The background from $e^+e^-\to~\qqbar$
($q=u, d, s$ and $c$ quarks) continuum events is represented by a linear
function.


\section{Background suppression}

   An important issue for this analysis is the suppression of the large
combinatorial background, which is dominated by $\qqbar$ continuum events.
We suppress this background using
variables that characterize the event topology. A more detailed description
of the background suppression technique can be found in Ref.~\cite{garmash}.

\begin{figure}[!t]
  \centering
  \includegraphics[width=0.49\textwidth]{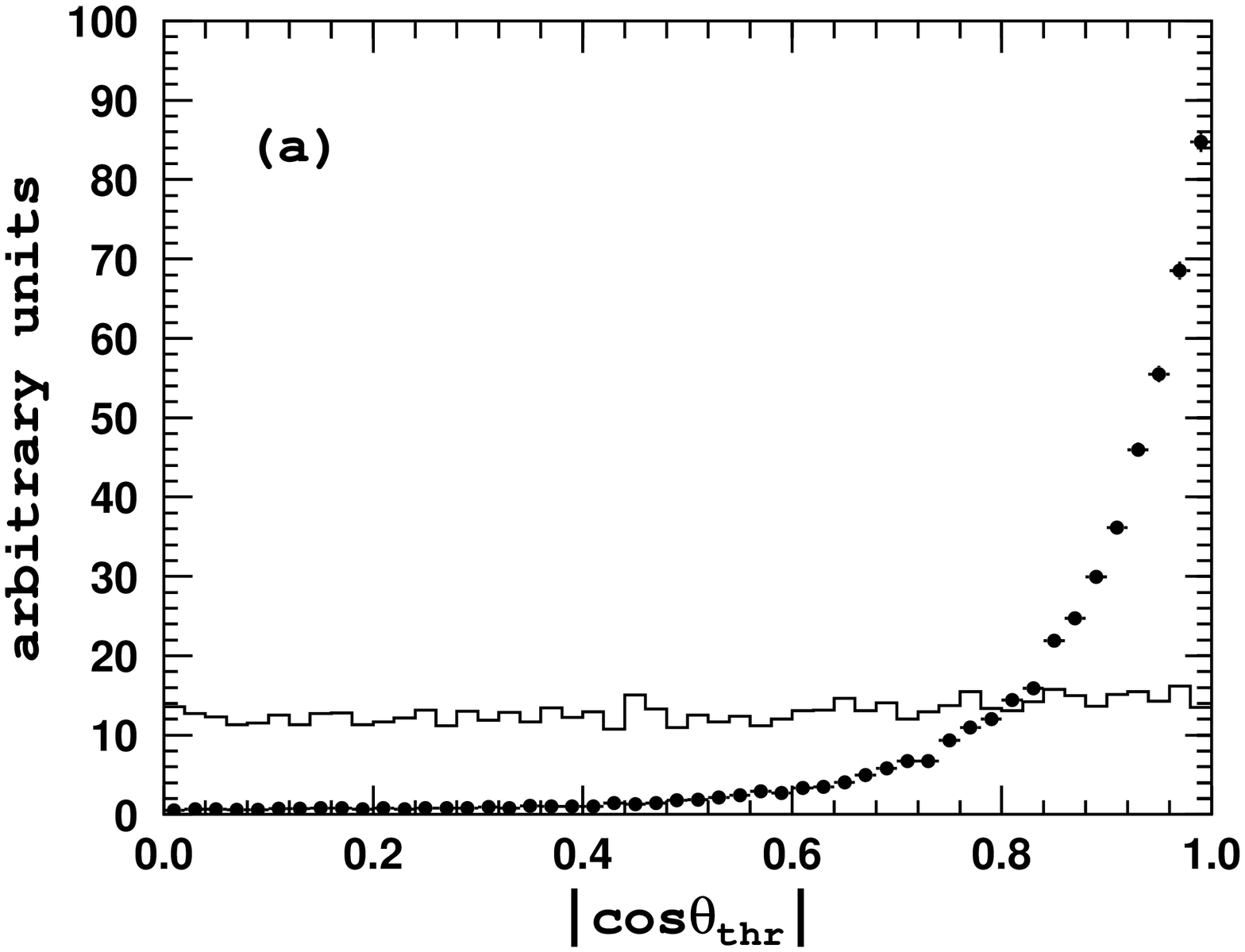}  \hfill
  \includegraphics[width=0.49\textwidth]{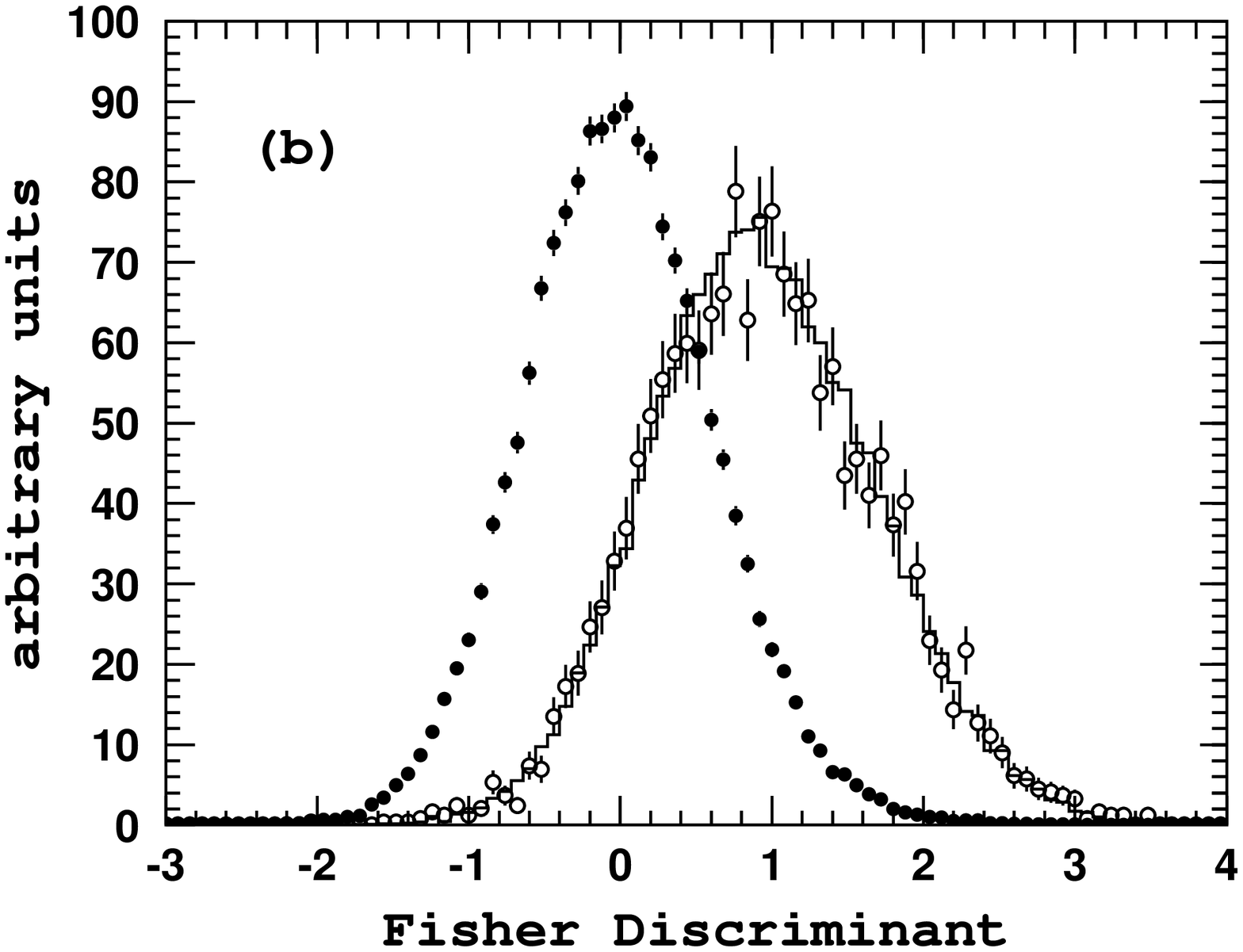}  \\
  \caption{Distribution of (a) $|\cos\theta_{\rm thr}|$ and (b) Fisher
           discriminant for the $\bckppos$ signal MC events (histogram),
           $B^+\to\bar D^0 \pi^+$ events in data (open circles) and $\qqbar$
           background in below-resonance data (filled circles).}
  \label{fig:shape}
\end{figure}

 Since the two $B$ mesons produced from an $\Upsilon (4S)$ decay are nearly at
rest in the c.m.\ frame, their decay products are uncorrelated and events tend
to be spherical. In contrast, hadrons from continuum $\qqbar$ events tend to
exhibit a two-jet structure. We use $\theta_{\rm thr}$, which is the angle
between the thrust axis of the $B$ candidate and that of the rest of the event,
to discriminate between the two cases. The distribution of
$|\cos\theta_{\rm thr}|$, shown in Fig.~\ref{fig:shape}(a), is strongly peaked
near $|\cos\theta_{\rm thr}|=1.0$
for $\qqbar$ events and is nearly flat for $\bbbar$ events. We require
$|\cos\theta_{\rm thr}|<0.80$ for all three-body final states; this eliminates
about 83\% of the continuum background and retains 79\% of the signal events.

   After imposing the $\cos\theta_{\rm thr}$ requirement, the remaining
$\qqbar$ and $B\bar{B}$ events still have some differences in topology that
are exploited for further continuum suppression. A Fisher
discriminant~\cite{fisher} is formed from 11 variables: nine variables
that characterize the angular distribution of the momentum flow in the event
with respect to the $B$ candidate thrust axis~\cite{VCal}, the angle of
the $B$ candidate thrust axis with respect to the beam axis, and the angle
between the $B$ candidate momentum and the beam axis. The discriminant,
$\cal{F}$, is the linear combination of the input variables that maximizes
the separation between signal and background. The coefficients are determined
using below-resonance data and a large set of $\bckppos$ signal MC events.
We use the same set of coefficients for all three-body final states.
Figure~\ref{fig:shape}(b) compares the $\cal{F}$ distributions for the
$\bckppos$ signal MC events, $B^+\to\bar D^0 \pi^+$ events in data and and
$\qqbar$ background events in below-resonance data. The separation between the
mean values of the discriminant for the signal and background is approximately
1.3 times the signal width. For all $K\pi\pi$ and $KK\pi$ final states we
impose a requirement ${\cal{F}}>0.8$ on the Fisher discriminant variable that
rejects about 90\% of the remaining continuum background and retains 54\% of
the signal. The continuum background in the three-kaon final states is much
smaller and a looser requirement ${\cal{F}}>0$ is imposed to retain the
efficiency. This requirement rejects 53\% of the remaining $\qqbar$ background
and retains 89\% of the signal.

We also consider backgrounds that come from other $B$ decays. We subdivide
this background into two types. The first type is the background from decays
that are dominantly $b\to c$ tree transition. The description of
these decays is taken from an updated version of the CLEO group event
generator~\cite{qqcleo}. The other potential source of background is
charmless $B$ decays that proceed via $b\to s(d)$ penguins or $b\to u$ tree
transitions. We studied a large set of potentially serious backgrounds from
two-, three-, and four-body final states. Since the background from other $B$
decays is substantially mode dependent, we give a detailed description of
this type of background for each final state in the following sections.

\section{$\bcnkcnpp$ }

We find that the dominant background to the $\kcnpp$ final states from other
$B$ decays is due to the $B\to D\pi$, $D\to K\pi$ decays and due to
$B^+\to J/\psi(\psi(2S)) K^+$ decays, where $J/\psi(\psi(2S))\to\mu^+\mu^-$.
To exclude the $B\to D\pi$ events, we apply the requirement on the $K\pi$
invariant mass $|M(K\pi)-M_D|>0.10$~GeV/$c^2$, where $M_D$ is the world average
value for mass of the $D^{0(+)}$ meson~\cite{PDG}. To suppress the background
due to $\pi/K$ misidentification in the $\kppos$ final state, we also exclude
candidates if the invariant mass of any pair of oppositely charged tracks from
the $B$ candidate is consistent with the $\bar{D}^0\to K^+\pi^-$ hypothesis
within 15~MeV/$c^2$ ($\sim 2.5\sigma$), independently of the particle
identification information. Modes with a $J/\psi(\psi(2S))$ in the final state
contribute due to muon-pion misidentification; the contribution from the
$J/\psi(\psi(2S))\to e^+e^-$ submode is found to be negligible (less than
0.5\%) after the electron veto requirement. We exclude $J/\psi(\psi(2S))$
signals by requiring $|M(h^+h^-)-M_{J/\psi}|>0.07$~GeV/$c^2$ and
$|M(h^+h^-)-M_{\psi(2S)}|>0.05$~GeV/$c^2$, where $h^+$ and $h^-$ are pion
candidates~\cite{psiveto}. Finally, we also reject candidates if the $\pipi$
invariant mass is within 50~MeV/$c^2$ of the world average $\chi_{c0}$
mass~\cite{PDG}. The most significant background from charmless $B$ decays is
found to originate from the $B^{+(0)}\to \eta'K^{+(0)}$ followed by
$\eta'\to \pi^+\pi^-\gamma$ decays. Another contribution comes from the
$B^+\to\rho^0\pi^+$ final state, where one of the final state pions is
misidentified as a kaon. There is also a background from two-body charmless
$B\to K\pi$ decays. Although this background is shifted by about 0.2~GeV
from the $\de$ signal region, it is important to take it into account to
estimate correctly the background from $\qqbar$ continuum events.

\begin{figure}[!t]
  \centering
  \includegraphics[width=0.49\textwidth]{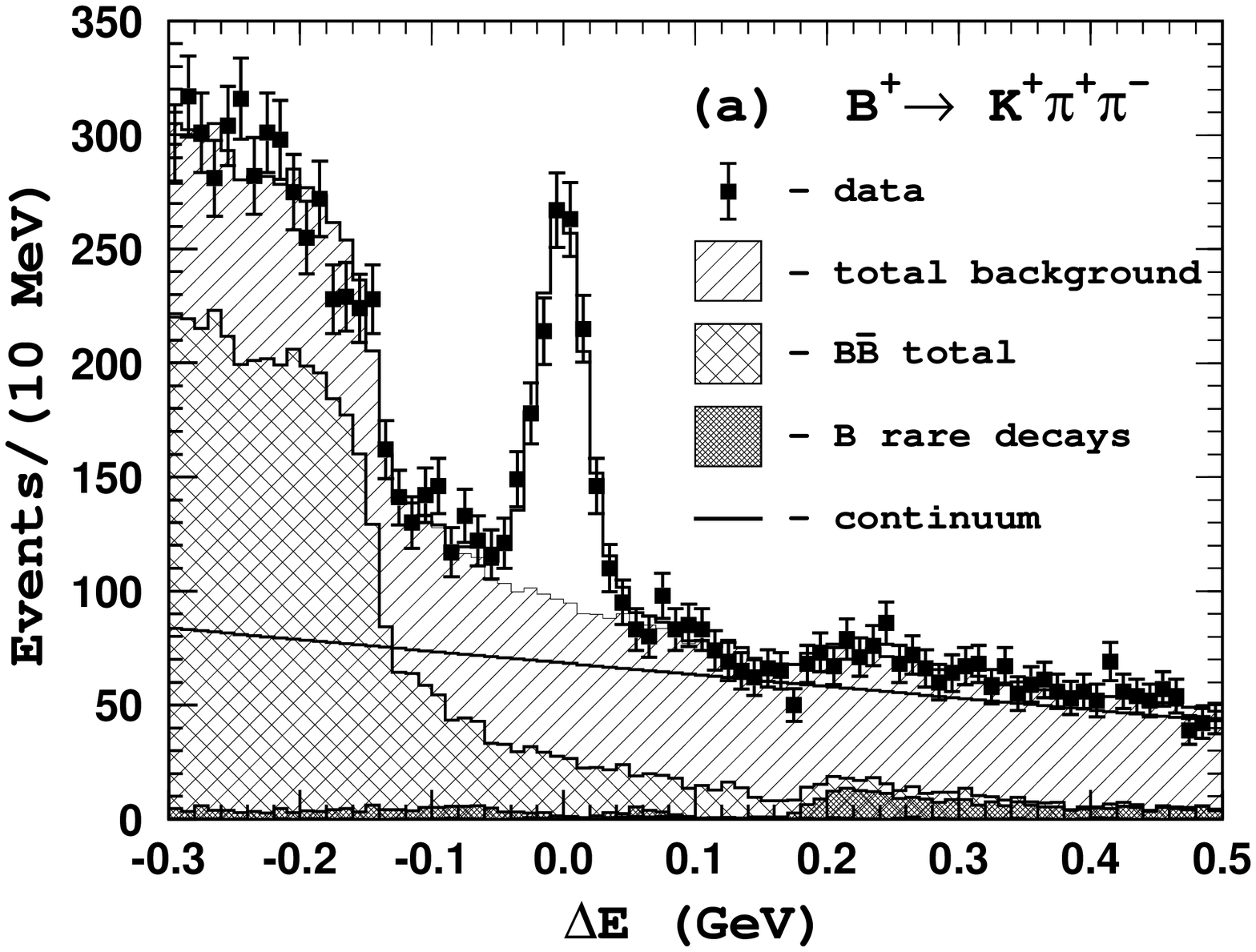}  \hfill
  \includegraphics[width=0.49\textwidth]{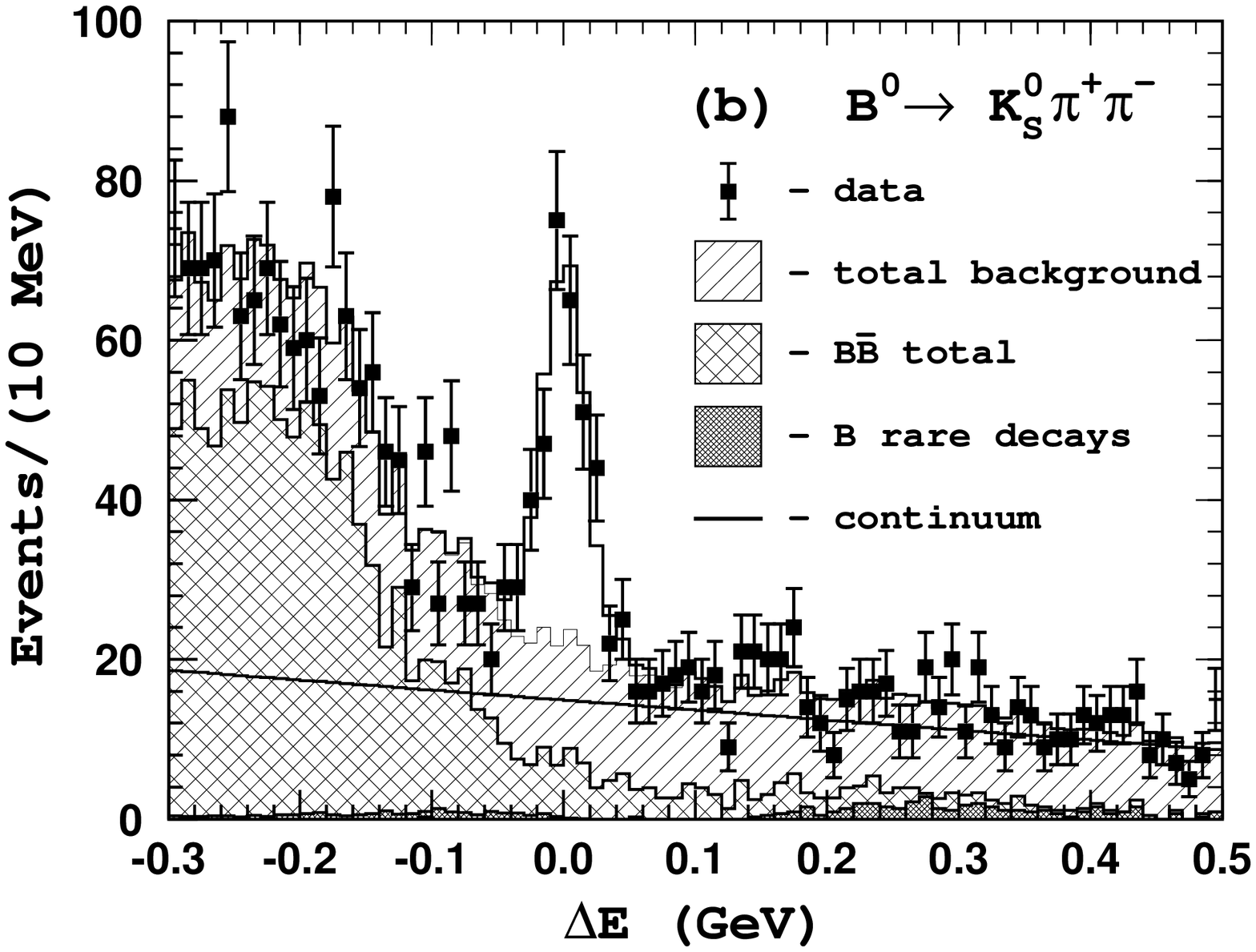}  \\
  \caption{$\de$ distributions for
           (a) $B^+\to K^+\pi^+\pi^-$ and
           (b) $B^0\to K^0\pi^+\pi^-$.
           Points with error bars represent data, the open histogram is the
           result of the fit, and the hatched histogram shows the total
           background level. The straight line indicates the
           $\qqbar$ continuum background contribution.}
  \label{fig:kpp_de}
\end{figure}

The $\de$ distributions for selected $\bcnkcnpp$ candidates that pass all
the selection requirements are shown in Fig.~\ref{fig:kpp_de}. A significant
enhancement in  the $B$ signal region is observed for both final states.
Results of the fits are shown as open histograms in Fig.~\ref{fig:kpp_de} and
hatched histograms represent the background. While fitting the data we fix
the shape of the $\bbbar$ background and allow its normalization to float. For
background from charmless $B$ decays we fix the normalization as well. Both the
normalization and the slope of the $\qqbar$ background are floated during the
fit. There is a large increase in the
level of the $\bbbar$ related background in the $\de<-0.15$~GeV region that
is mainly due to $B\to D\pi$, $D\to K\pi\pi$ decays. This decay mode produces
the same final state as the studied process plus one extra pion that is not
included in the energy difference calculation. The decay $B\to D\pi$,
$D\to K\mu\nu_\mu$ also contributes due to muon-pion misidentification. The
shape of this background is well described by MC simulation. The signal yields
from fits are given in Table~\ref{tab:defit}.

\section{$B \to KKK$}

The dominant background from other $B$ decays to the three-kaon final states is
from $B\to Dh$ decays, where $h$ stands for a charged pion or kaon. To suppress
this background, we reject events where the two-particle invariant mass is
consistent within 15~MeV/$c^2$ ($\sim$2.5$\sigma$) with $D^0\to K^+K^-$ for the
$\kkk$ final state and with $D^+\to\bar{K}^0K^+$ for the $\knkk$ final state.
To suppress the background due to $\pi/K$ misidentification in the $\kkk$
final state, we also exclude candidates if the invariant mass of any pair of
oppositely charged tracks from the $B$ candidate is consistent with
$D^0\to K^-\pi^+$ within 15~MeV/$c^2$ ($\sim 2.5\sigma$), independently of the
particle identification information. For the $\knkk$ final state we exclude
candidates if the $K^0_S h^\pm$ invariant mass is consistent with
$D^+\to\bar{K}^0\pi^+$ within 15~MeV/$c^2$. We also reject events with a
$\kpkm$ invariant mass that is consistent with $\chi_{c0}\to K^+K^-$ within
50~MeV/$c^2$. We do not find any charmless $B$ decay modes that produce
a significant background to the three-kaon final states. The feed-across
between $\kcnpp$ and $\kcnkk$ final states is also found to be negligible.

\begin{figure}[!t]
  \includegraphics[width=0.49\textwidth]{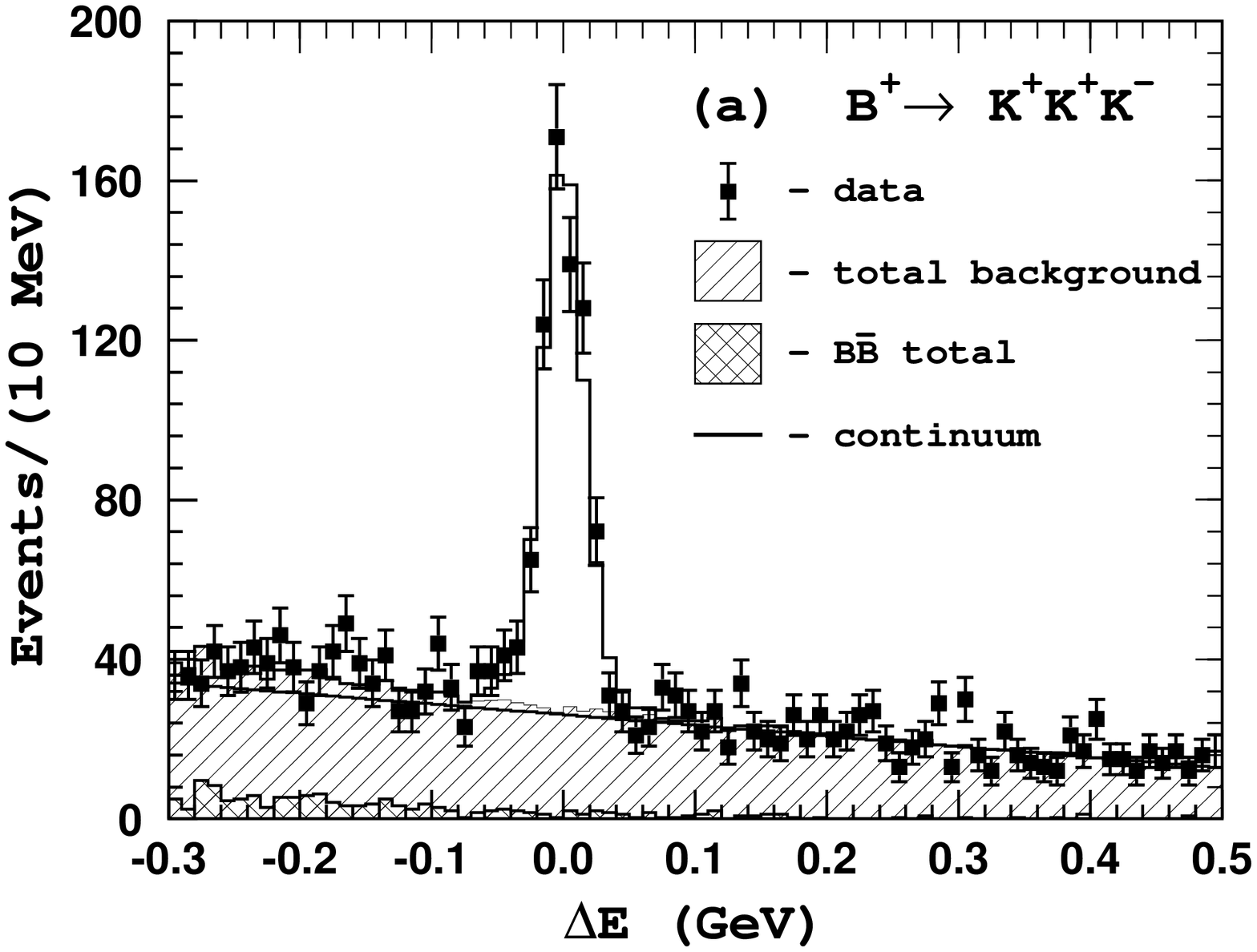} \hfill
  \includegraphics[width=0.49\textwidth]{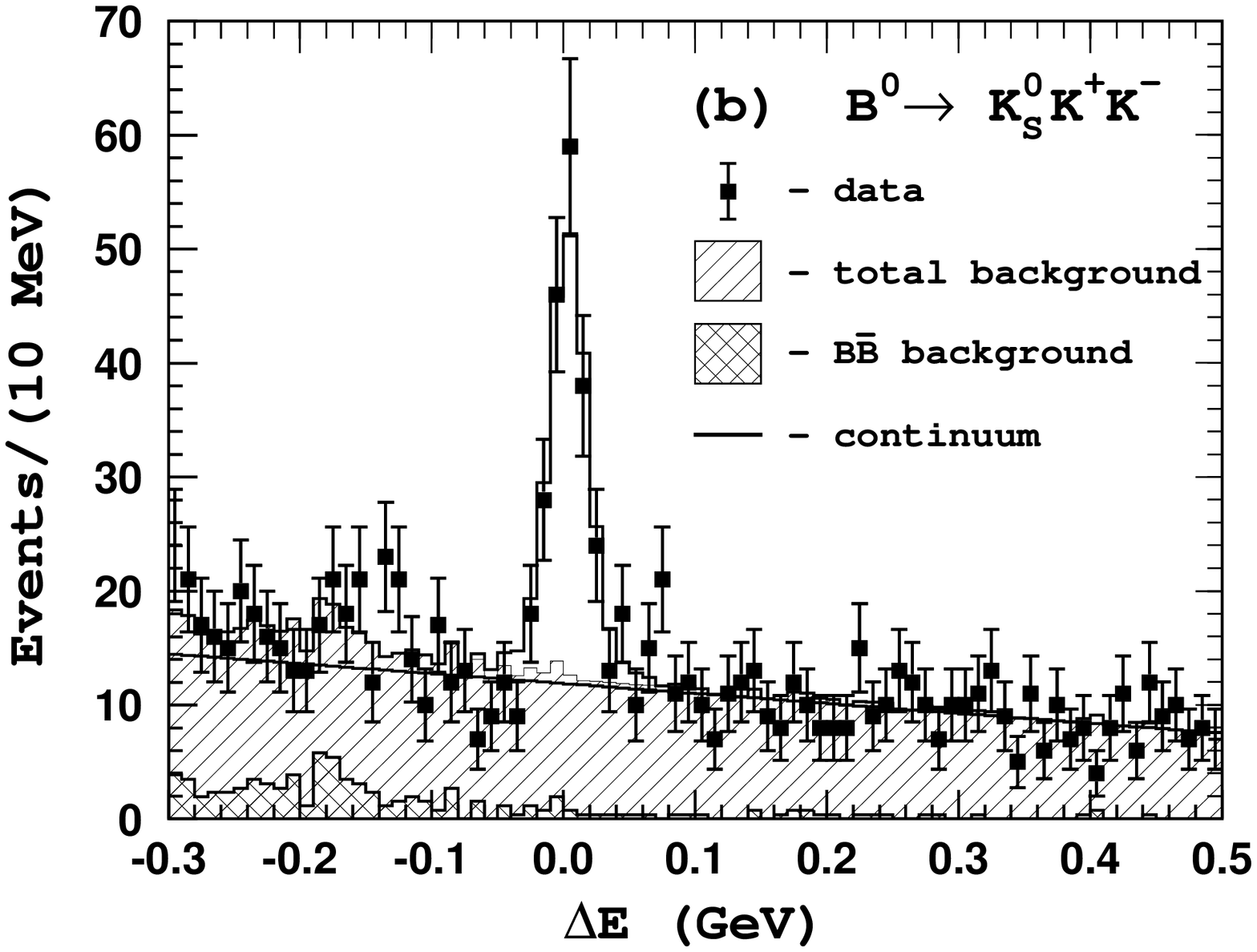}\\
  \includegraphics[width=0.49\textwidth]{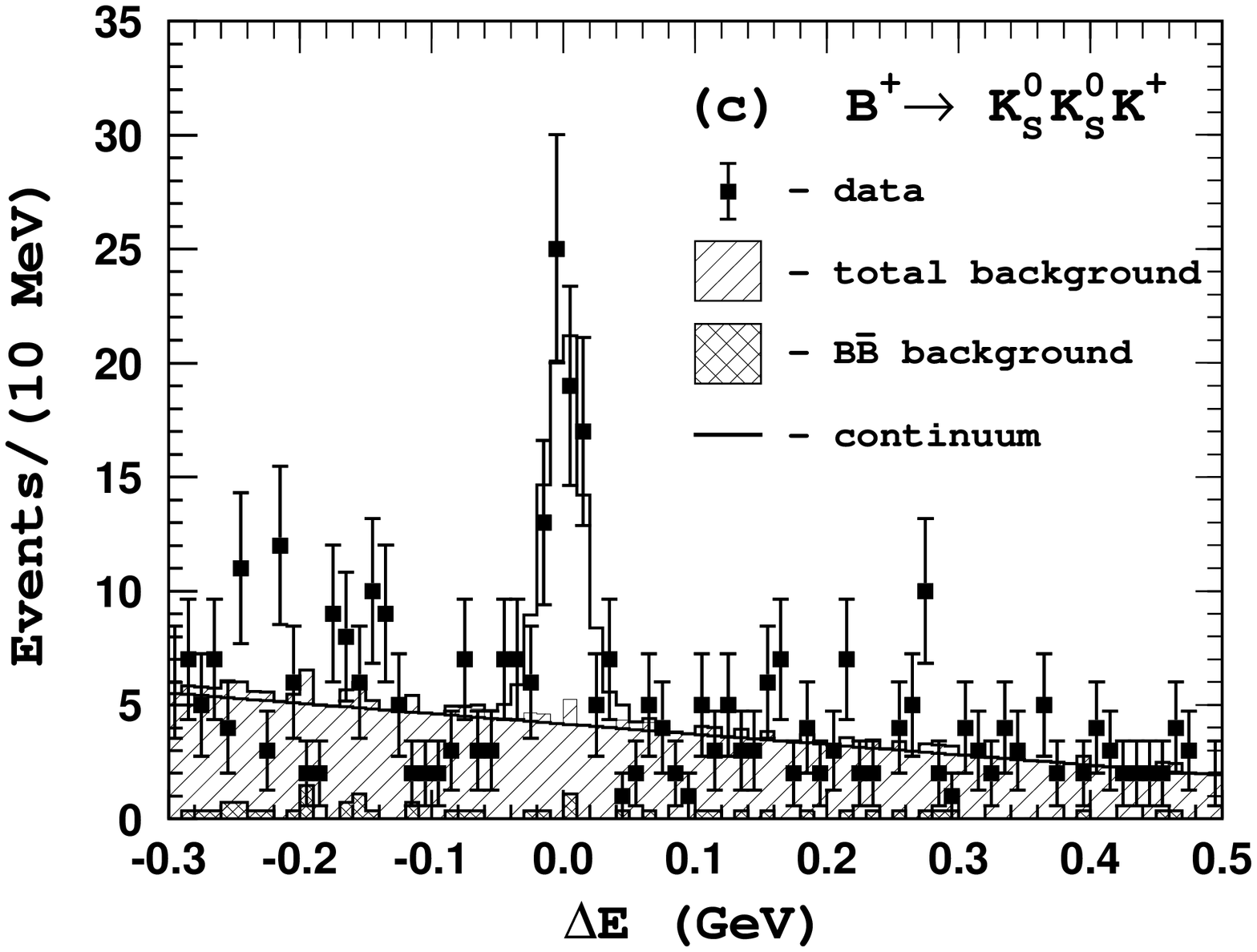} \hfill
  \includegraphics[width=0.49\textwidth]{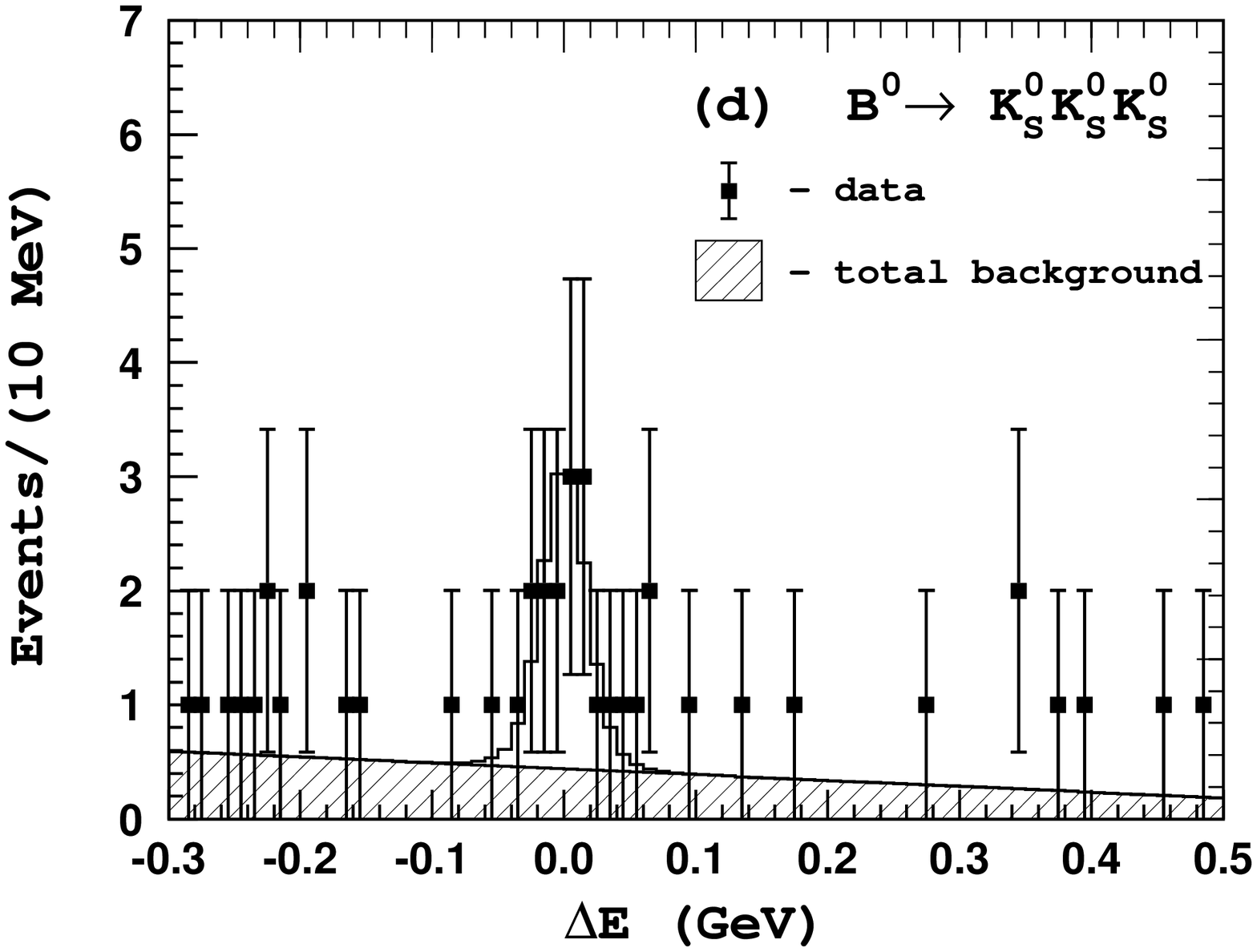}\\
  \caption{$\de$ distributions for $B\to KKK$ three-body final states.
           Points with error bars are data; the open histogram is the fit
           result; the hatched histogram is the background. The straight line
           shows the $\qqbar$ continuum background contribution.}
  \label{fig:kkkDE}
\end{figure}

  The $\de$ distributions for all three-kaon final states are shown in
Fig.~\ref{fig:kkkDE}, where data (points with errors) are shown along with the
expected background (hatched histograms).  The $\bbbar$ background in $KKK$
final states is much smaller than that in the $\kcnpp$ final states and has no
prominent structures. The results of fits to the $\de$ distributions are
summarized in Table~\ref{tab:defit}. While fitting the data for all three-kaon
final states we fix not only the shape but also the normalization of the
$\bbbar$ related background. The statistical significance of the $\bnksksks$
signal, in terms of the number of standard deviations is 4.3$\sigma$. It is
calculated as $\sqrt{-2\ln({\cal{L}}_0/{\cal{L}}_{\rm max})}$, where
${\cal{L}}_{\rm max}$ and ${\cal{L}}_{0}$ denote the maximum likelihood with
the nominal signal yield and with the signal yield fixed at zero, respectively.
The significance of the signal in all other three-kaon final states exceeds
10$\sigma$.

\section{$\bcnkcnkp, ~\ksksp, ~\kkpss$ and $\kppss$}

The signals in the $\kcnkp$ and $\ksksp$ channels are, in general, expected
to be much smaller since the dominant contributions to these final states are
expected to come from the $b\to u$ tree and $b\to dg$ penguin transitions,
while the $\bcnkcnpp$ and $B\to KKK$ decay channels are dominantly $b\to sg$
penguin transitions. As for the $\bckkpss$ and $\bckppss$ decays, the SM
predicts exceedingly small branching fractions for these decays (as mentioned
above).

\begin{figure}[pthb]
\centering
  \includegraphics[width=0.49\textwidth]{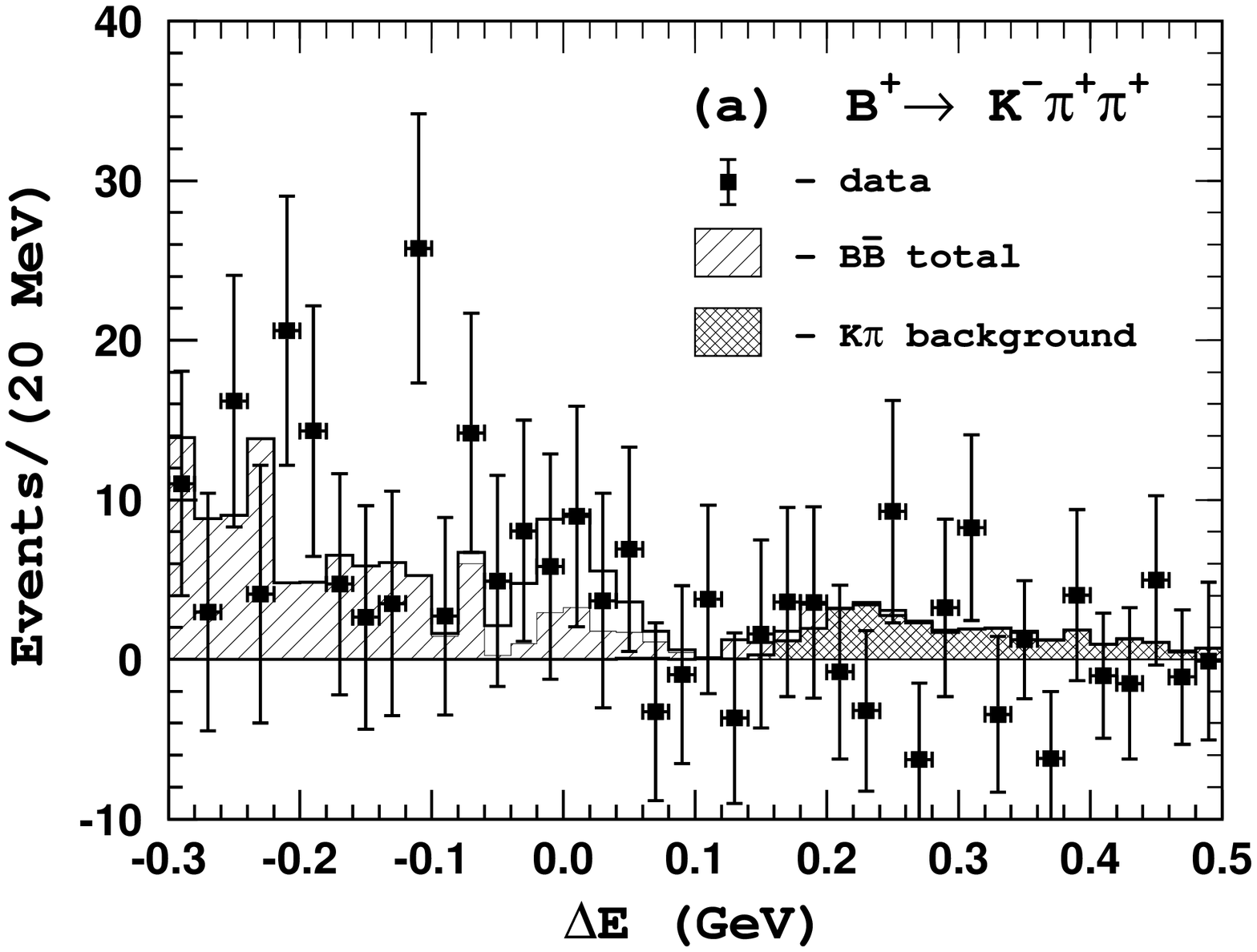} \hfill
  \includegraphics[width=0.49\textwidth]{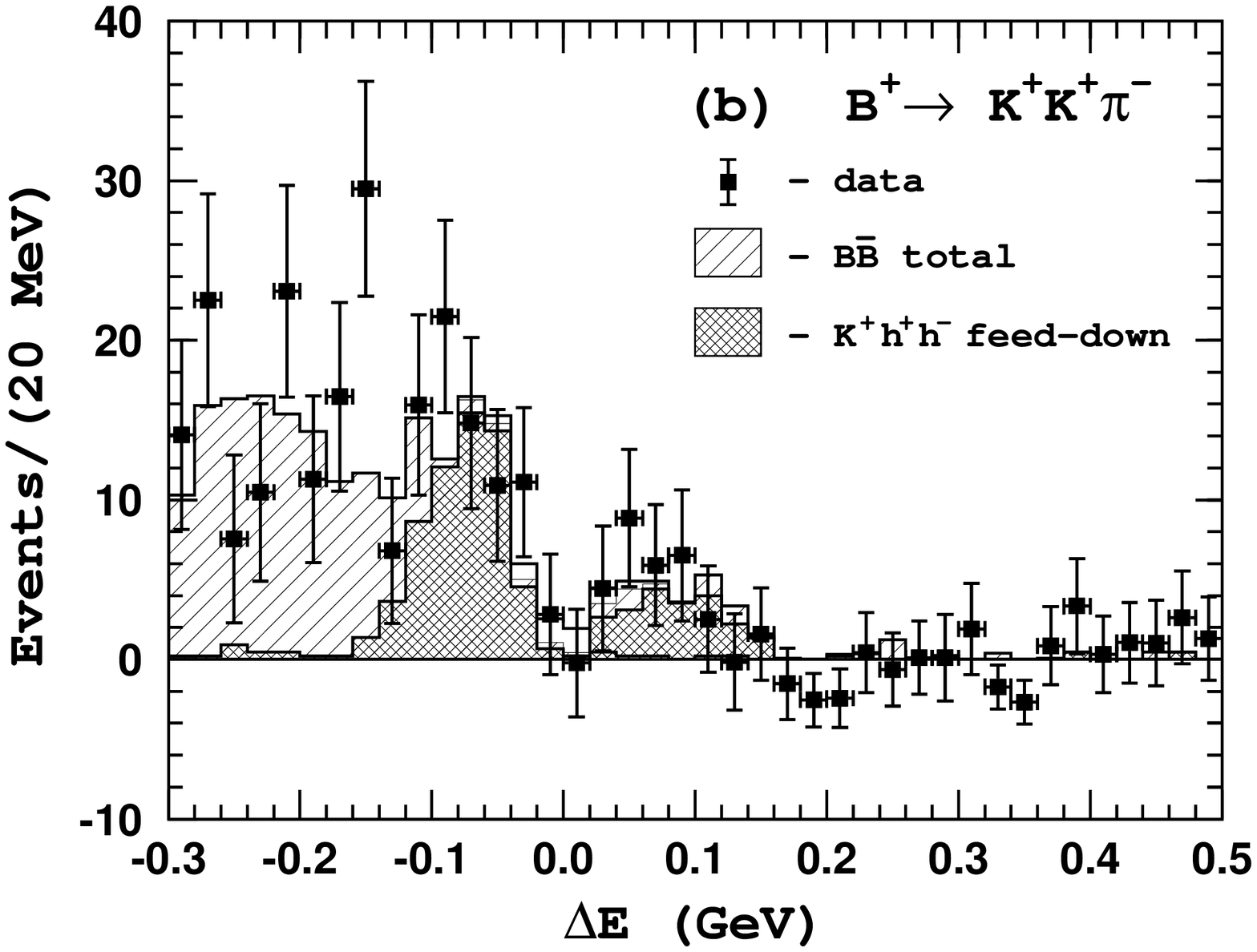} \\
  \includegraphics[width=0.49\textwidth]{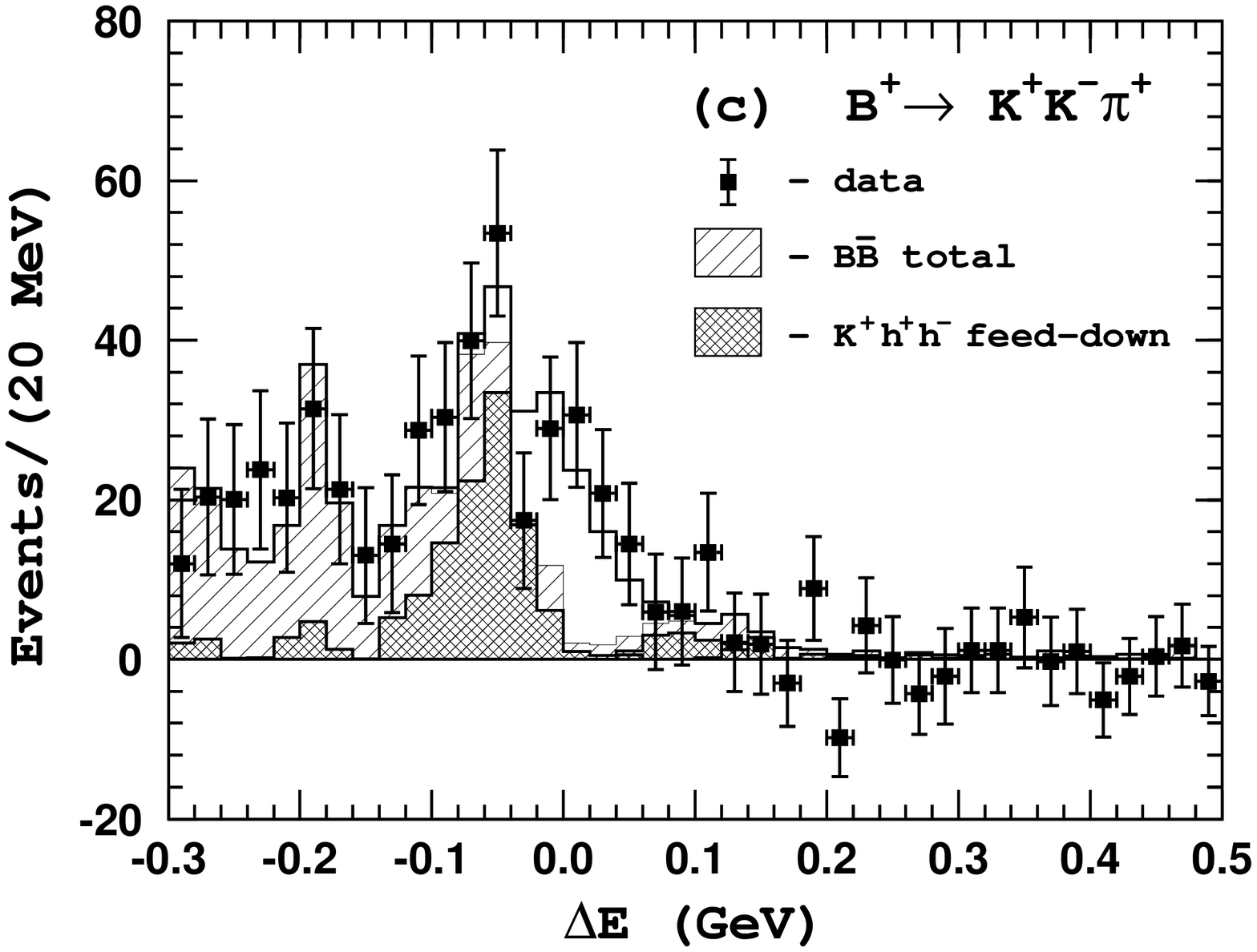} \hfill
  \includegraphics[width=0.49\textwidth]{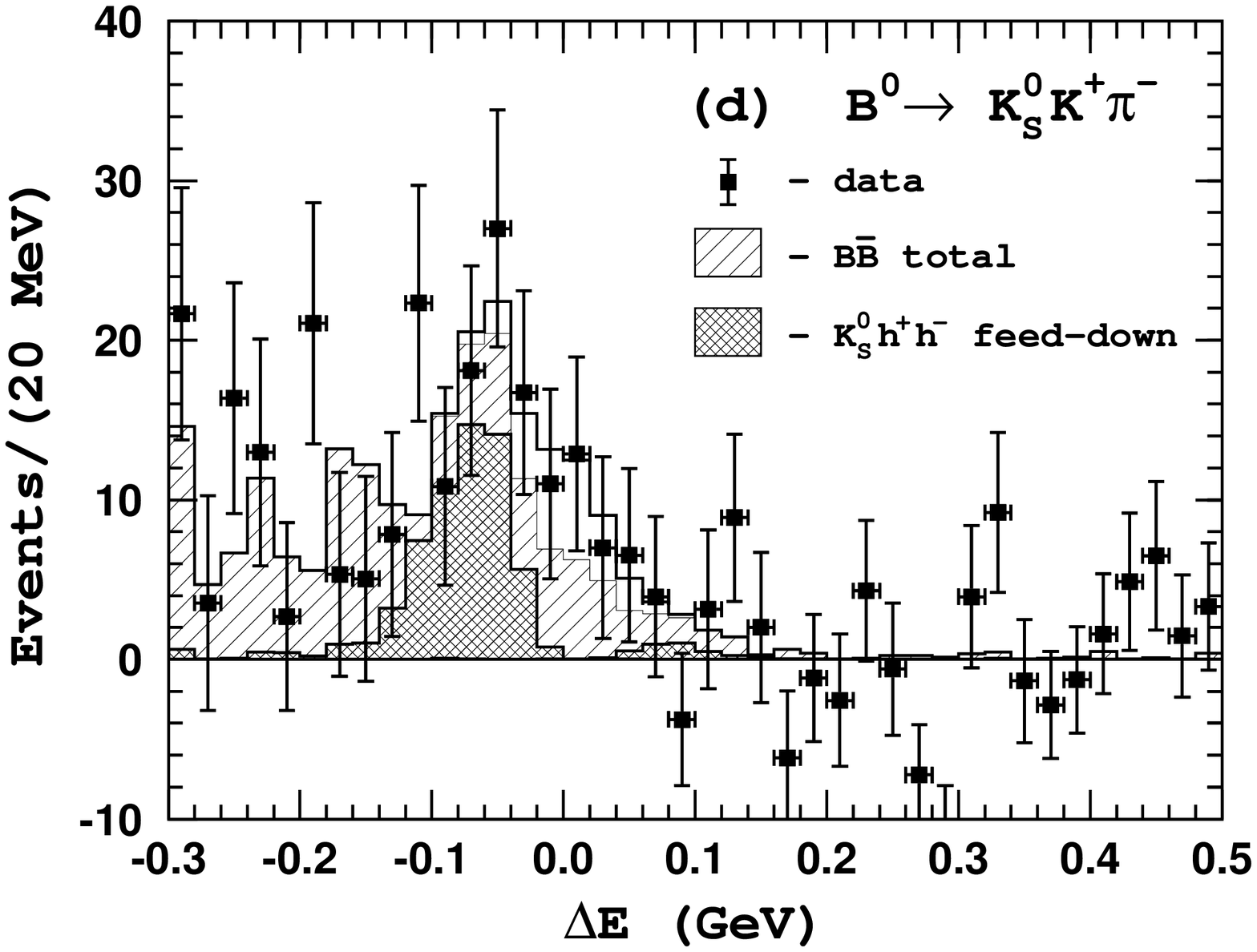} \\
  \includegraphics[width=0.49\textwidth]{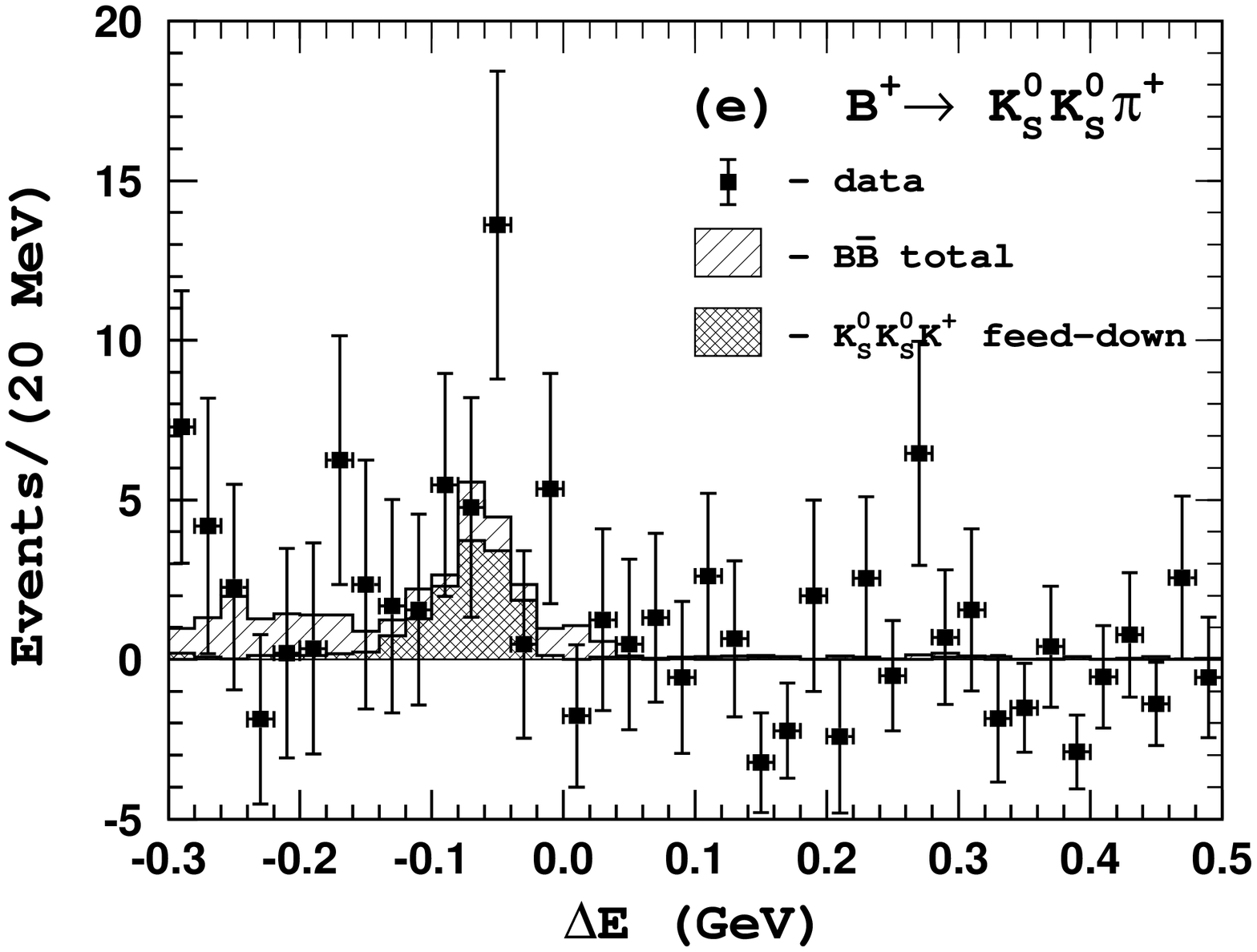}
\hspace*{0.504\textwidth}
  \centering
  \caption{Results of the fit (open histogram) to the $\mb$ distributions
           in $\de$ bins.
           The data (points with error bars) are compared with the $B\bar{B}$
           MC expectation (hatched histogram). The feed-down from
           $KKK$ and $K\pi\pi$ final states is shown by filled histograms.}
  \label{fig:kkp_de}
\end{figure}

The background from other $B$ meson decays to these final states is mainly due
to $B\to Dh$ decays, where $h$ represents a charged pion or kaon. We suppress
this background by rejecting events where the two-particle invariant mass is
consistent within 15~MeV/$c^2$ ($\sim$2.5$\sigma$) with
$D^0\to K^+K^-$, $D^0\to K^-\pi^+$, $D^+\to\bar{K}^0\pi^+$ or
$D^+\to\bar{K}^0K^+$. Because of the small expected signal to background ratio
for these final states, a different technique for signal yield extraction is
used. We subdivide the $\de$ region into 20~MeV bins and determine the signal
yield in each bin from the fit to the corresponding $\mb$ spectrum. The $\mb$
signal shape is parameterized by a Gaussian function. The width of the $\mb$
signal is primarily due to the c.m.\ energy spread and is expected to be the
same for each decay channel; in the fit we fix it at the value
$\sigma_{\mb}=3.0$~MeV/$c^2$ determined from the $B^+\to\bar{D^0}\pi^+$,
$\bar{D^0}\to K^+\pi^-$ events in the same data sample. The background shape
is parameterized with the empirical function
$f(\mb)\propto x\sqrt{1-x^2}\exp[-\xi(1-x^2)]$, where $x = \mb/E_{\rm beam}$
and $\xi$ is a parameter~\cite{ArgusF}. We fix the $\xi$ value from a study of
data taken below the $\Upsilon(4S)$ resonance. The signal yield from the $\mb$
fit as a function of $\de$ is used in the subsequent analysis. Since the
$\qqbar$ background does not peak in the $\mb$ distribution, this technique
allows an effective subtraction of the $\qqbar$ background. In contrast, the
$B\bar{B}$ background can easily produce a signal-like distribution  in the
$\mb$ variable. The same procedure is applied to the $B\bar{B}$ MC events to
determine the background shape and level. The resulting
continuum-background-subtracted $\de$ distributions are fit with the signal
yield as the only free parameter. Figure~\ref{fig:kkp_de} shows the results of
the fit, along with the expected contributions from generic $B\bar{B}$ decays
and the feed-down due to particle misidentification from the $B\to K\pi\pi$ and
$B\to KKK$ decay modes. For the $\kppss$ final state, shown in
Fig.~\ref{fig:kkp_de}(a), the only background from charmless $B$ decays is due
to
$\bar{B}^0\to K^-\pi^+$ two-body decay. For the $\kkpos$ final state, shown in
Fig.~\ref{fig:kkp_de}(c), we observe an excess of events in the signal region
with statistical significance of $4\sigma$. For all other final states
presented in Fig.~\ref{fig:kkp_de}, the experimental points are consistent with
the background expectations.


\section{Branching fraction Calculation}

   To determine branching fractions, we normalize our results to the observed
$B^+\to\bar{D}^0\pi^+$, $\bar{D}^0\to K^+\pi^-$ and
$B^0\to D^-\pi^+$, $D^-\to K^0\pi^-$ signals. This reduces the
systematic errors associated with the charged track reconstruction efficiency,
particle identification efficiency, the event shape variables and uncertainty
due to the possible nonuniform data taking conditions during the experiment.
We calculate the branching fraction for $B$ meson decay to a particular final
state $f$ via the relation
\begin{equation}
 {\cal{B}}(B\to f) =
   \frac{N_f}{N_{D\pi}}\frac{\varepsilon_{D\pi}}{\varepsilon_{f}}\times
   {\cal{B}}(B\to D\pi){\cal{B}}(D\to K\pi),
\end{equation}
where $N_f$ and $N_{D\pi}$ are the number of reconstructed signal events for
the final state $f$ and that for the $D\pi$ reference process, and
$\varepsilon_{f}$ and $\varepsilon_{D\pi}$ are the corresponding reconstruction
efficiencies determined from MC. The $B\to D\pi$ and $D\to K\pi$ branching
fractions are the world average values from the PDG~\cite{PDG}.

\begin{figure}[!t]
  \includegraphics[width=0.49\textwidth]{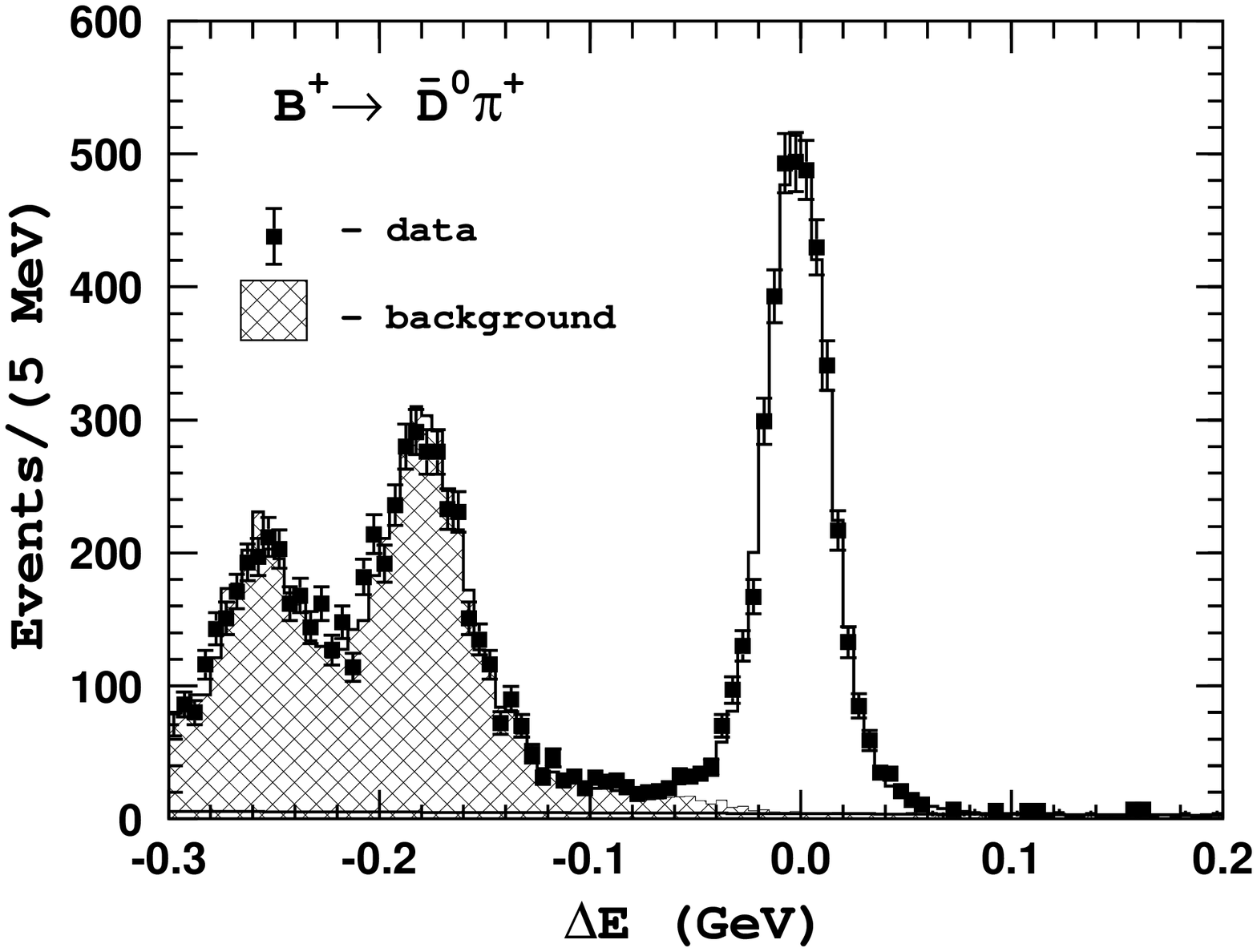} \hfill
  \includegraphics[width=0.49\textwidth]{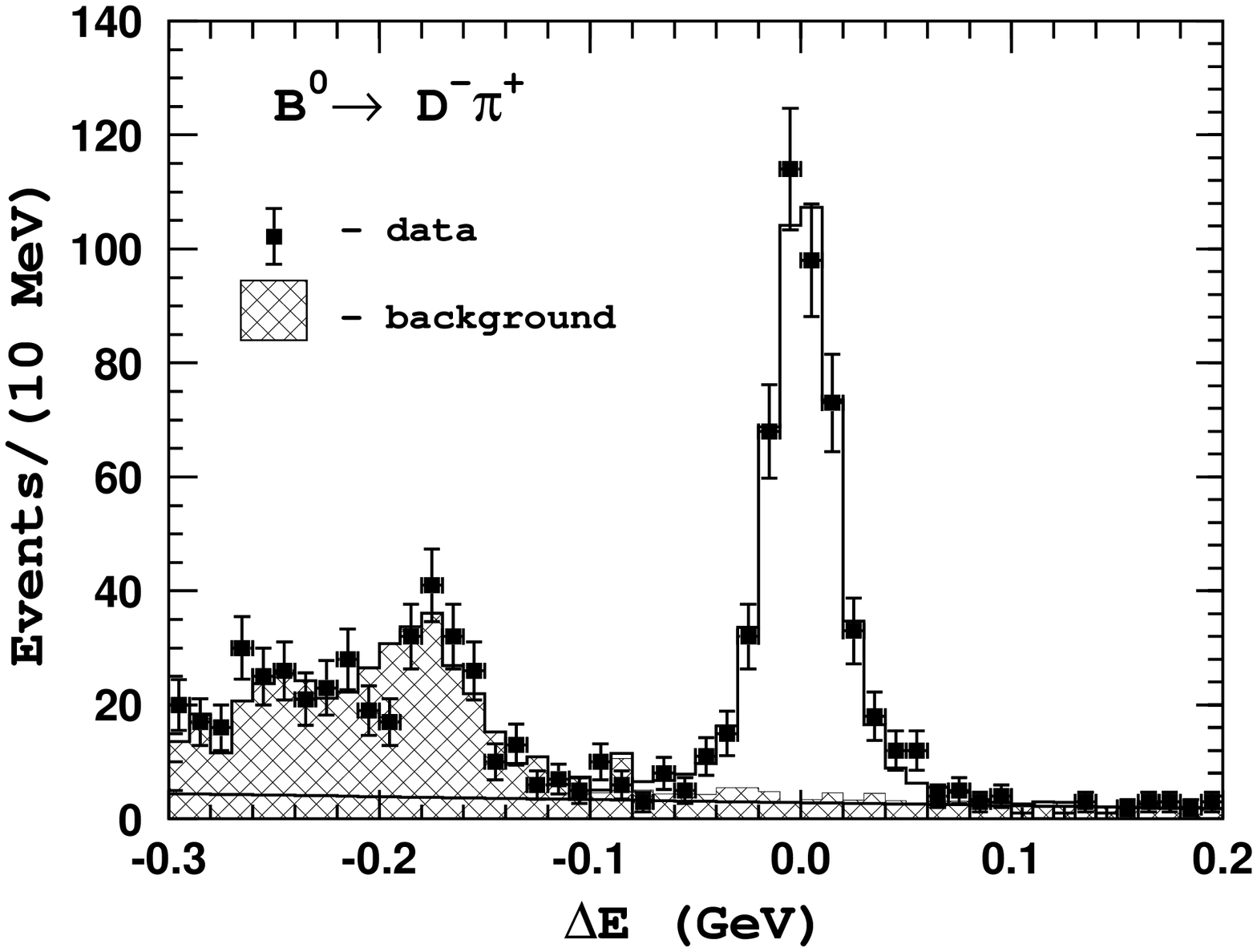}\\
  \caption{$\de$ distributions for $B^+\to\bar{D}^0\pi^+$,
           $\bar{D}^0\to K^+\pi^-$ (left) and  $B^0\to D^-\pi^+$,
           $D^-\to K^0\pi^-$ (right) events. Points with error bars represent
           data with ${\cal{F}}>0$, the open histogram is the fit
           result and the hatched histogram is the background. The straight
           line shows the $\qqbar$ continuum background contribution.}
  \label{fig:dpiDE}
\end{figure}

To select $B\to D\pi$ events, we require that the $K\pi$ invariant mass be
consistent within 3 $\sigma$ with the $D$ meson nominal mass~\cite{PDG}.
The $\de$ distributions for the reference processes $B^+\to\bar{D}^0\pi^+$,
$\bar{D}^0\to K^+\pi^-$ and $B^0\to D^-\pi^+$, $D^-\to K^0\pi^-$ are shown in
Fig.~\ref{fig:dpiDE}. The results of the fits are given in
Table~\ref{tab:defit}. Since, in the analysis of $K\pi\pi$ ($KK\pi$) and $KKK$
final states, different requirements on the Fisher discriminant are applied, we
determine the $B\to D\pi$ signal yields for these two cases.

The results of the three-body branching fraction measurements are presented in
Table~\ref{tab:defit}. To determine the reconstruction efficiencies for $\kkk$
and $\knkk$ final states, we use a simple model~\cite{garmash} that takes into
account the non-uniform distribution of signal events over the Dalitz plot. The
three-body signal in this model is parameterized by a $\phi K$ intermediate
state and a $f_X K$ state, where $f_X$ is a hypothetical wide scalar state.
Another model is used for the $\kcnpp$ final states to account for the
non-uniform distribution of signal events over the Dalitz plot and determine
the reconstruction efficiencies~\cite{garmash}. The three-body $\kcnpp$ signal
in this model is parameterized by the following set of quasi-two-body
intermediate states: $\rho^0 K$, $f_0(980) K$, $K^*(892)\pi$, $K^*_0(1430)\pi$
and $f_0(1370) \pi$. For the $\kppss$, $\ksksk$, $\ksksks$ and all $KK\pi$
final states, the reconstruction efficiency is determined from MC simulated
events that are generated with a uniform (phase space) distribution over the
Dalitz plot. For the $\bckkpos$ decay the statistical significance of the
signal barely exceeds 4~$\sigma$ and as a final result we set a 90\% confidence
level upper limit on its branching fraction, though the central value is also
given in Table~\ref{tab:defit}. Since we do not observe a statistically
significant signal in any of the $\kppss$, $\kkpss$, $\knkp$ and $\ksksp$ final
states, we place 90\% confidence level upper limits on their branching
fractions~\cite{ul-calc}. These limits are given in Table~\ref{tab:defit}.
To take into account the systematic uncertainty, we reduce the reconstruction
efficiency by one standard deviation of the overall systematic error.

\begin{table}[t]
  \caption{Summary of results for $B$ meson decays to three-body charmless
           hadronic final states. 
Table lists the three-body decay modes, the corresponding $\de$ resolution
for the core ($\sigma_1$) and the tail ($\sigma_2$) Gaussian functions,
the fraction of the signal in the core Gaussian, three-body reconstruction
efficiency, the signal yield extracted from the $\de$ fit and branching
fractions.
The branching fractions and 90\% confidence
           level (CL) upper limits (UL) are quoted in units of $10^{-6}$.
           For the modes with neutral kaons the quoted
           reconstruction efficiencies include the intermediate branching
           fractions. Two values for the $B\to D\pi$ efficiency and
           signal yield correspond to ${\cal{F}}>0.8$ ($>0$) requirement on
           the Fisher discriminant.}
  \medskip
  \label{tab:defit}
  \begin{tabular}{lccccc} \hline \hline
\hspace*{0mm}  Three-body   \hspace*{-1mm} &
\hspace*{1mm}  $\de$ resolution $\sigma_1/\sigma_2$   \hspace*{1mm}  &
\hspace*{1mm}  fraction     \hspace*{1mm}  &
\hspace*{1mm}  Efficiency   \hspace*{1mm}  &
\hspace*{1mm}  Signal Yield \hspace*{1mm}  &
\hspace*{1mm}  $\cal B$ or 90\% CL UL     \hspace*{0mm}  \\
 ~~~~mode  &  (MeV)  &        &     (\%)      &   (events)   & ($10^{-6}$)\\
\hline
 $\kppos$  & $17.3/35.0$ & $0.85$ & $21.5\pm0.48$ & $845\pm46$   & $\bfkppos$ \\
 $\knpp$   & $15.2/40.0$ & $0.85$ & $5.85\pm0.11$ & $209\pm21$   & $\bfknpp$  \\
\hline
 $\kkk$    & $14.5/40.0$ & $0.85$ & $23.5\pm0.50$ & $565\pm30$   & $\bfkkk$   \\
 $\knkk$   & $14.0/40.0$ & $0.85$ & $7.20\pm0.17$ & $149\pm15$   & $\bfknkk$  \\
 $\ksksk$  & $14.3/40.0$ & $0.85$ & $6.78\pm0.19$ & $66.5\pm9.3$ & $\bfksksk$ \\
 $\ksksks$ & $14.7/40.0$ & $0.85$ & $3.98\pm0.17$ & $12.2^{+4.5}_{-3.8}$
                                          & $\bfksksks$  \\
\hline
 $\kkpos$  & $15.5/40.0$ & $0.85$ & $13.8\pm0.31$ &  $94\pm23~(<130)$   & $\bfkkpos~(\ulkkpos)$ \\
 $\knkp$   & $14.7/40.0$ & $0.85$ & $4.53\pm0.16$ &   $27\pm17~(<55)$   & $\ulknkp$  \\
 $\ksksp$  & $15.0/40.0$ & $0.85$ & $5.31\pm0.15$ & $-1.8\pm7.7~(<11)$  & $\ulksksp$ \\
\hline
 $\kppss$  & $17.0/40.0$ & $0.85$ & $17.0\pm0.37$ &  $21\pm18~(<51)$    & $\ulkppss$ \\
 $\kkpss$  & $15.5/40.0$ & $0.85$ & $14.2\pm0.30$ & $6.5\pm9.6~(<22.3)$ & $\ulkkpss$ \\
\hline
 $\bar{D}^0\pi^+$ & $14.5/30.0$ & $0.80$ & $17.8\pm0.43~(28.8\pm0.57)$ & $2470\pm51~(4000\pm66)$ & $-$ \\
 $D^-\pi^+$       & $14.9/40.0$ & $0.80$ & $5.10\pm0.11~(8.24\pm0.14)$ & $ 300\pm16~( 451\pm25)$ & $-$ \\
\hline \hline
  \end{tabular}
\end{table}

The dominant sources of systematic error are listed in Table~\ref{khh_syst}.
For the $\bcnkcnpp$ and $\bcnkcnkk$ final states, we estimate the systematic
uncertainty due to variations of reconstruction efficiency over the Dalitz plot
by varying the relative phases of quasi-two-body states in the range from $0$
to $2\pi$ and their amplitudes within $\pm20$\%. For all other final states we
make a conservative estimate of this type of systematic uncertainty. For each
three-body final state the Dalitz plot is subdivided in 2~GeV$^2/c^4$ wide
slices in $M^2(hh)$ (for example $M^2(K^-\pi^+)$ for the $\kkpos$ final state),
and the reconstruction efficiency is determined in each bin. The same procedure
is then applied for the second Dalitz plot projection ($M^2(K^+K^-)$ for the
$\bckkpos$ decay). The maximal variation in the reconstruction efficiency for
all bins is taken as the systematic error. The uncertainty due to the particle
identification is estimated using high purity samples of kaons and pions from
the $D^0\to K^-\pi^+$ decays, where the $D^0$ flavor is tagged using
$D^{*+}\to D^0\pi^+$ decays. The systematic error due to uncertainty in the
$K^0_S$ reconstruction efficiency is estimated from the comparison of the
relative yields of inclusive $K^0_S$'s in off-resonance data and $\qqbar$ MC
with variation of the $K^0_S$ selection criteria. We estimate the uncertainty
due to the signal $\de$ shape parameterization by allowing the width of the
main Gaussian function to float and varying other parameters of the fitting
function within their errors. The uncertainty in the background
parameterization is estimated by varying the relative fractions of the
background components and parameters of the $\qqbar$ background shape function
within their errors. The overall systematic uncertainty for the three-body
branching fractions varies from 9\% to 18\%, as given in Table~\ref{khh_syst}.

\begin{table*}[t]
\centering
\caption{List of systematic errors (in percent) for the
         three-body branching fractions.}
\medskip
\label{khh_syst}
  \begin{tabular}{lccccccccccc}  \hline \hline
  \raisebox{7mm}{Source}
         &~~~\rotatebox{90}{$\kppos$}~~~
         &~~~\rotatebox{90}{$\knpp$}~~~
         &~~~\rotatebox{90}{$\kkk$}~~~
         &~~~\rotatebox{90}{$\knkk$}~~~
         &~~~\rotatebox{90}{$\ksksk$}~~~
         &~~~\rotatebox{90}{$\ksksks$}~~~
         &~~~\rotatebox{90}{$\kkpos$}~~~
         &~~~\rotatebox{90}{$\knkp$}~~~
         &~~~\rotatebox{90}{$\ksksp$}~~~
         &~~~\rotatebox{90}{$\kppss$}~~~
         &~~~\rotatebox{90}{$\kkpss$}~~~
 \\ \hline

$B\to D\pi$, $D\to K\pi$   &
  6.3 & 11.7 &  6.3 & 11.7 &  6.3 & 11.7 &  6.3 & 11.7 &  6.3 &  6.3 &  6.3 \\
    Branching fractions    &  \\ \hline

 Eff. non-uniformity       &
  4.3 &  3.5 &  2.2 &  3.6 &  5.9 &  7.2 &  5.3 &  5.7 &  6.1 &  6.8 &  5.3 \\
 over the Dalitz plot      &  \\ \hline

  Signal yield             & 
  4.9 &  3.6 &  2.1 &  5.6 &  4.8 &  7.6 &  6.8 &  $-$ &  $-$ &  $-$ & $-$  \\
  extraction               &  \\ \hline

      PID                  &
  $-$ &  $-$ &  4.0 &  4.0 &  2.0 &  2.0 &  2.0 &  2.0 &  $-$ &  $-$ &  2.0
                            \\ \hline
  $K^0_S$ reconstruction   &
  $-$ &  $-$ &  $-$ &  $-$ &  3.5 &  7.0 &  $-$ &  $-$ &  3.5 &  $-$ &  $-$
                            \\ \hline
     MC statistics       &
  3.3 &  2.9 &  2.9 &  2.9 &  3.4 &  4.6 &  3.3 &  4.1 &  3.7 &  3.3 &  3.2 \\
\hline \hline
Total                      &
  9.6 & 13.1 &  8.6 & 14.3 & 11.2 & 17.9 & 11.4 & 13.8 & 10.1 &  9.8 &  9.1 \\
\hline \hline
  \end{tabular}
\end{table*}


\section{Implication for CP violation study}

  An important check of the Standard Model would be provided by measurements
of the same CP-violating parameter in different weak interaction processes.
A good example is the comparison of the measurement of the coefficient of the
CP-violating $\sin(\Delta m_dt)$ term in the time dependent analysis of neutral
$B$ meson decays. In $B^0\to (c\bar{c})K^0$ decays (where $(c\bar{c})$ denotes
a charmonium state) this coefficient is $\sin2\phi_1$. Precise measurements of
$\sin2\phi_1$ (also referred to as $\sin2\beta$) have recently been reported by
the Belle and BaBar experiments~\cite{CP_phi1}. The best known candidates for
$b\to s$ penguin-dominated processes, where this quantity can be measured
independently, are $B^0\to\phi K^0$ and $B^0\to\eta' K^0$ decays. However, the
branching fractions for these decay modes are of order $10^{-6}-10^{-5}$
(including secondary branching fractions) and very large numbers of $B$
mesons are required to perform these measurements. This is especially true for
the $\phi K^0$ final state. The large signal observed in the three-body
$\bnkskk$ decay mode, where the $\phi K^0_S$ two-body intermediate state gives
a relatively small contribution~\cite{garmash,phik},
would significantly increase
the available statistics if these events could be used. Two complications
are involved:
(1) the $b\to u$ tree contributions may introduce an additional weak phase in
the $\bnkskk$ amplitude and complicate the interpretation of any observed CP
violation. The $b\to u$ contribution in $B^0\to\phi K^0_S$ is expected to be
negligible (since $\phi$ is almost a pure $s\bar{s}$ state), which is not
necessarily the case for the three-body $\kskk$ final state;
(2) In contrast to the $\phi K^0_S$ state, which has fixed CP-parity, the
CP-parity of the three-body $\kskk$ final state is not a priori known. If the
fractions of CP-even and CP-odd components are comparable, the use of the
$\kskk$ decay mode in a CP analysis will be complicated. Although in this case
an analysis of the proper time distribution would not be useful for CP
violation measurement, the analysis of the time evolution of the Dalitz plot
may still provide useful information on CP violation in the $\kskk$ final
state. In what follows we discuss the possibility of using the three-body
$\bnkskk$ decay mode for CP violation measurements~\cite{kskk-cp,GLNQ}.

\begin{figure}[t]
  \begin{minipage}[c]{1.1\textwidth}
  \hspace*{-1.1cm}\includegraphics[width=0.33\textwidth]{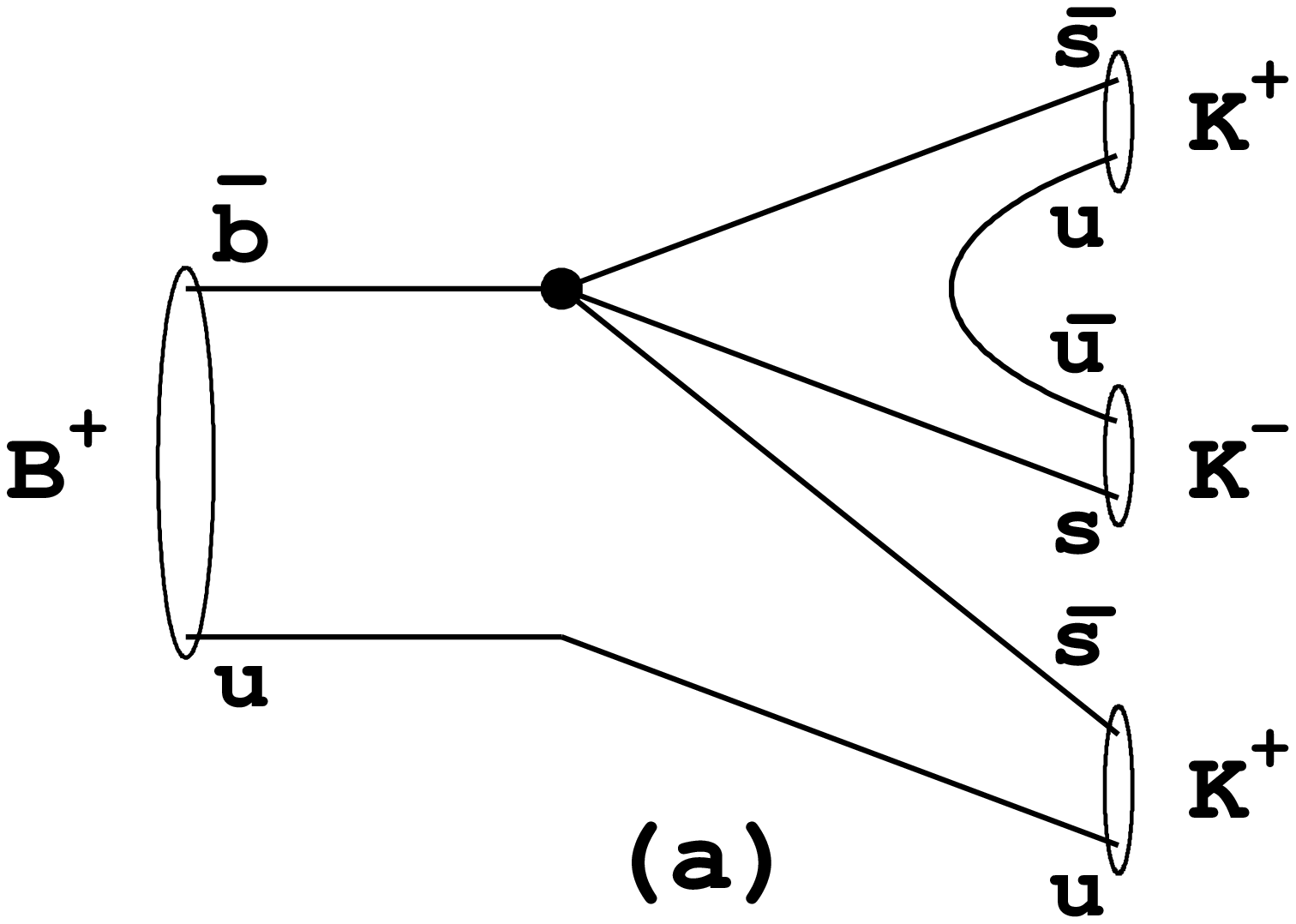}
  \hspace*{-0.3cm}\includegraphics[width=0.33\textwidth]{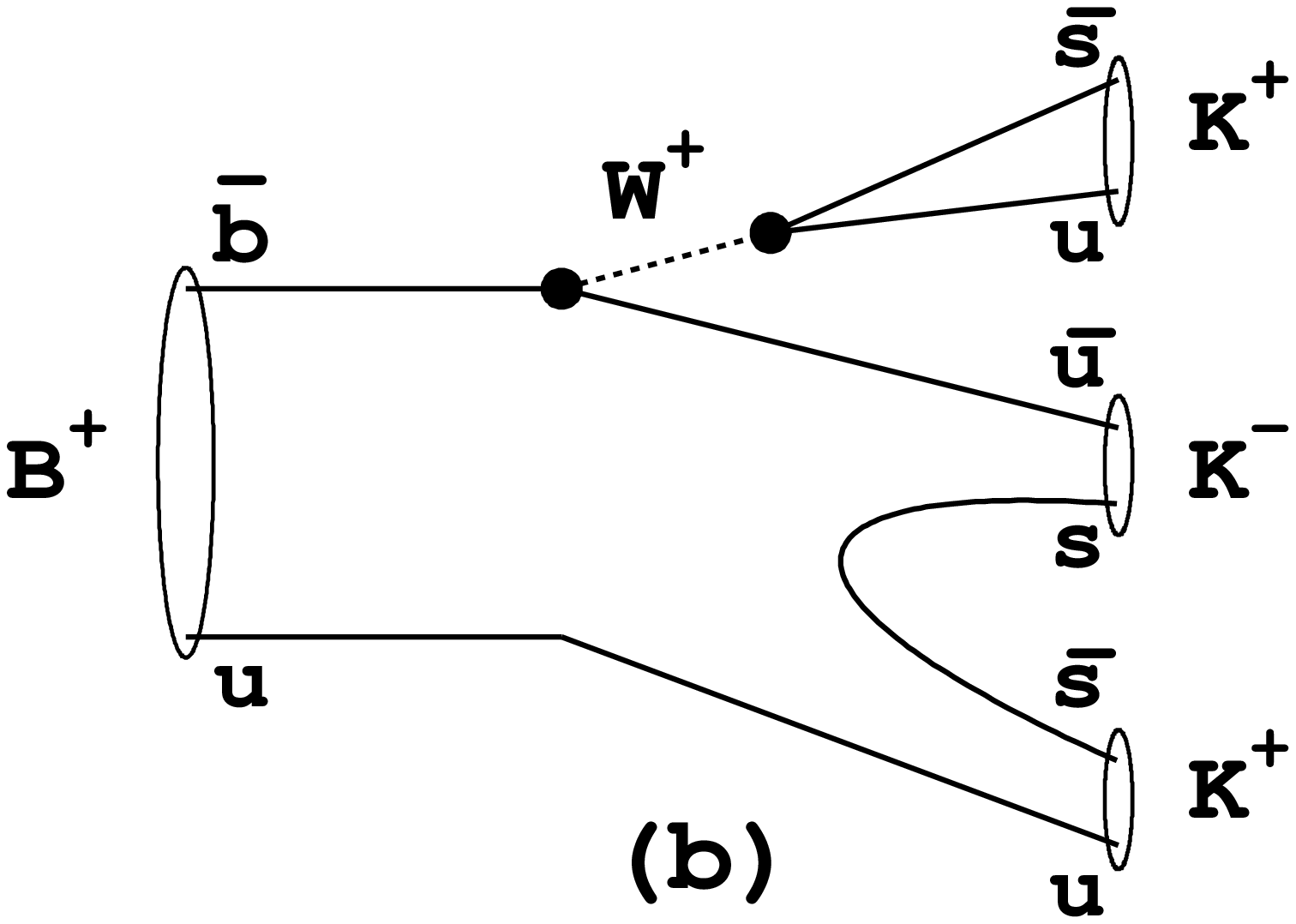}
  \hspace*{-0.6cm}\includegraphics[width=0.33\textwidth]{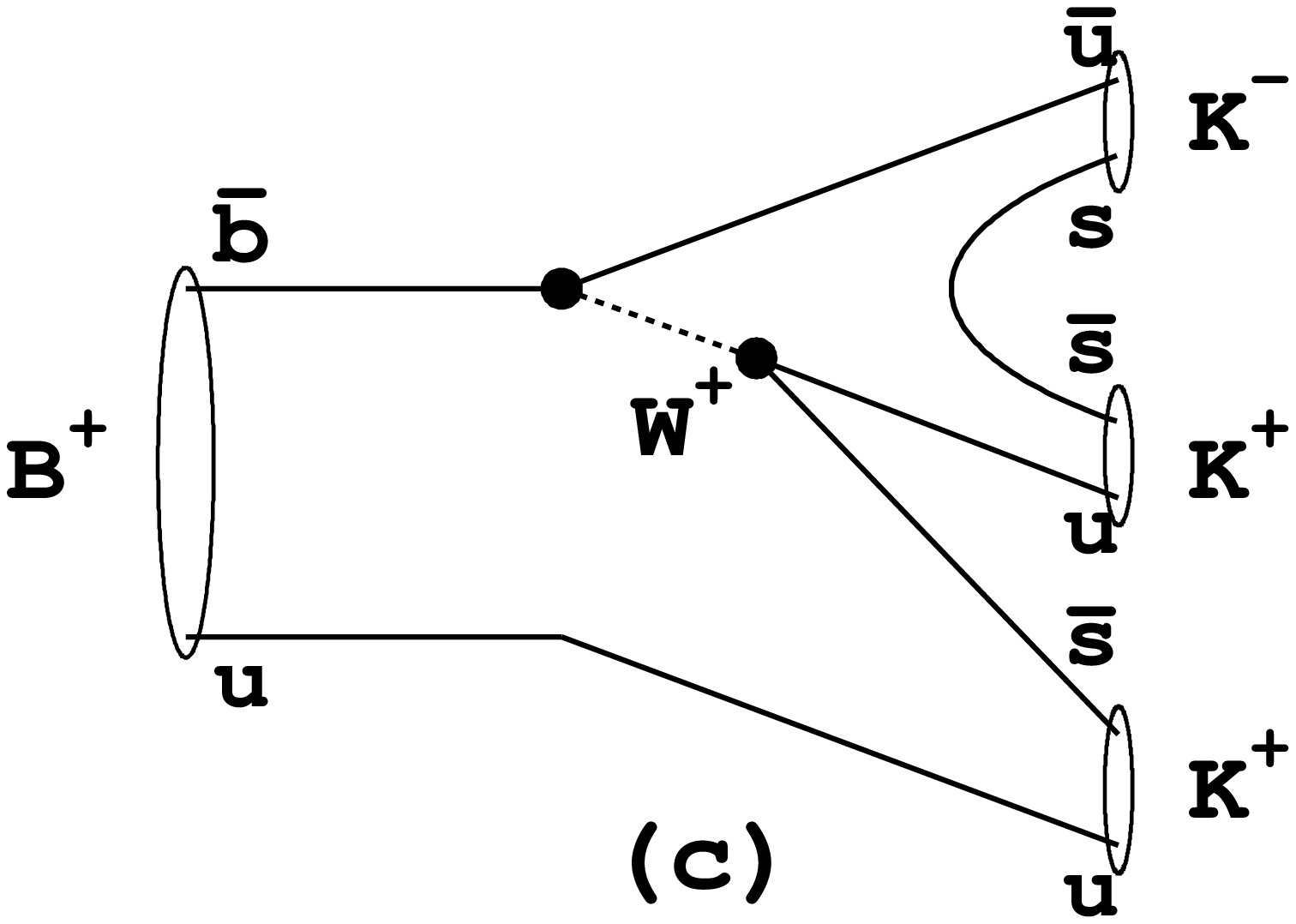}\\
  \hspace*{-1.1cm}\includegraphics[width=0.33\textwidth]{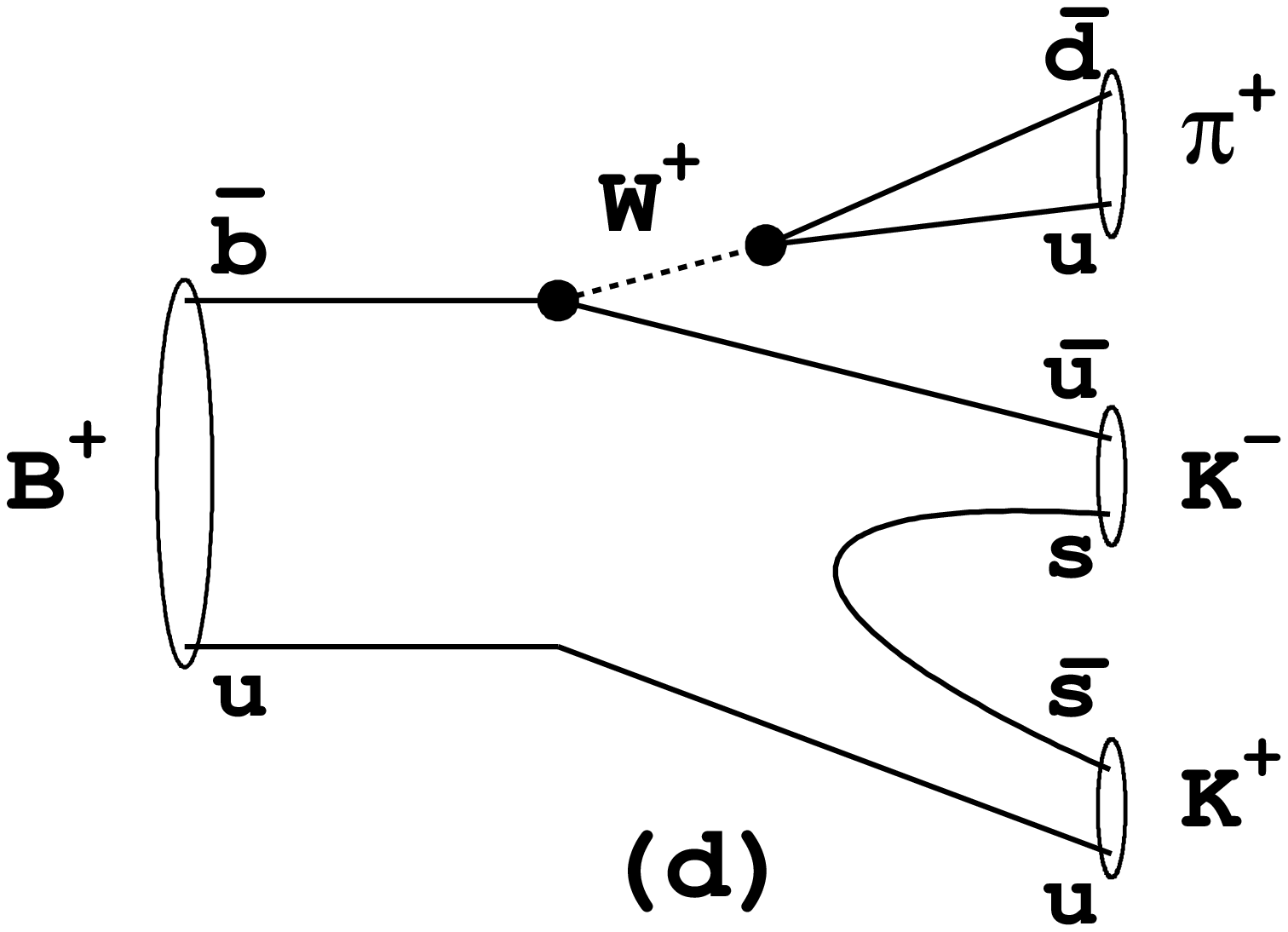}
  \hspace*{-0.3cm}\includegraphics[width=0.33\textwidth]{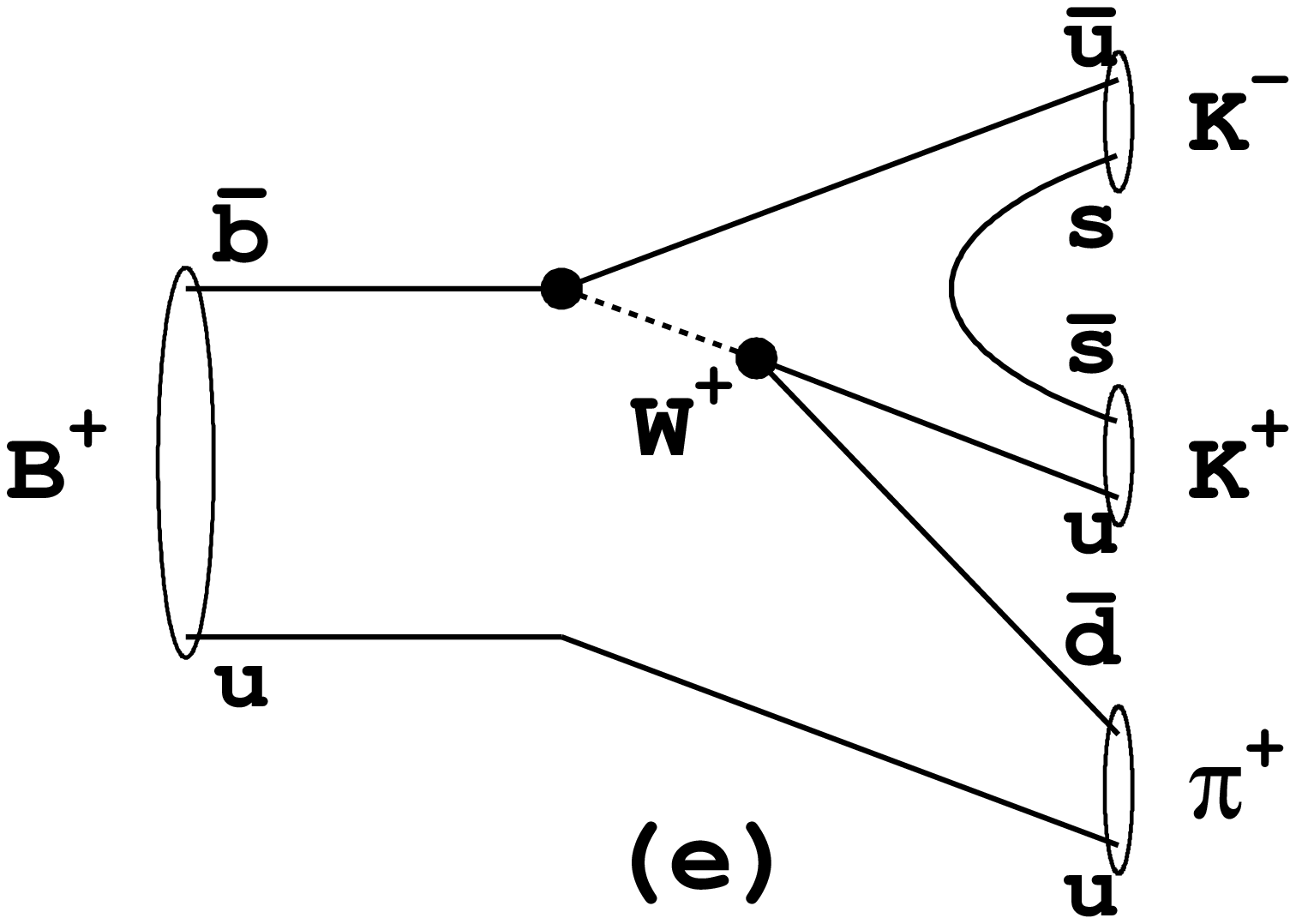}
  \hspace*{-0.6cm}\includegraphics[width=0.33\textwidth]{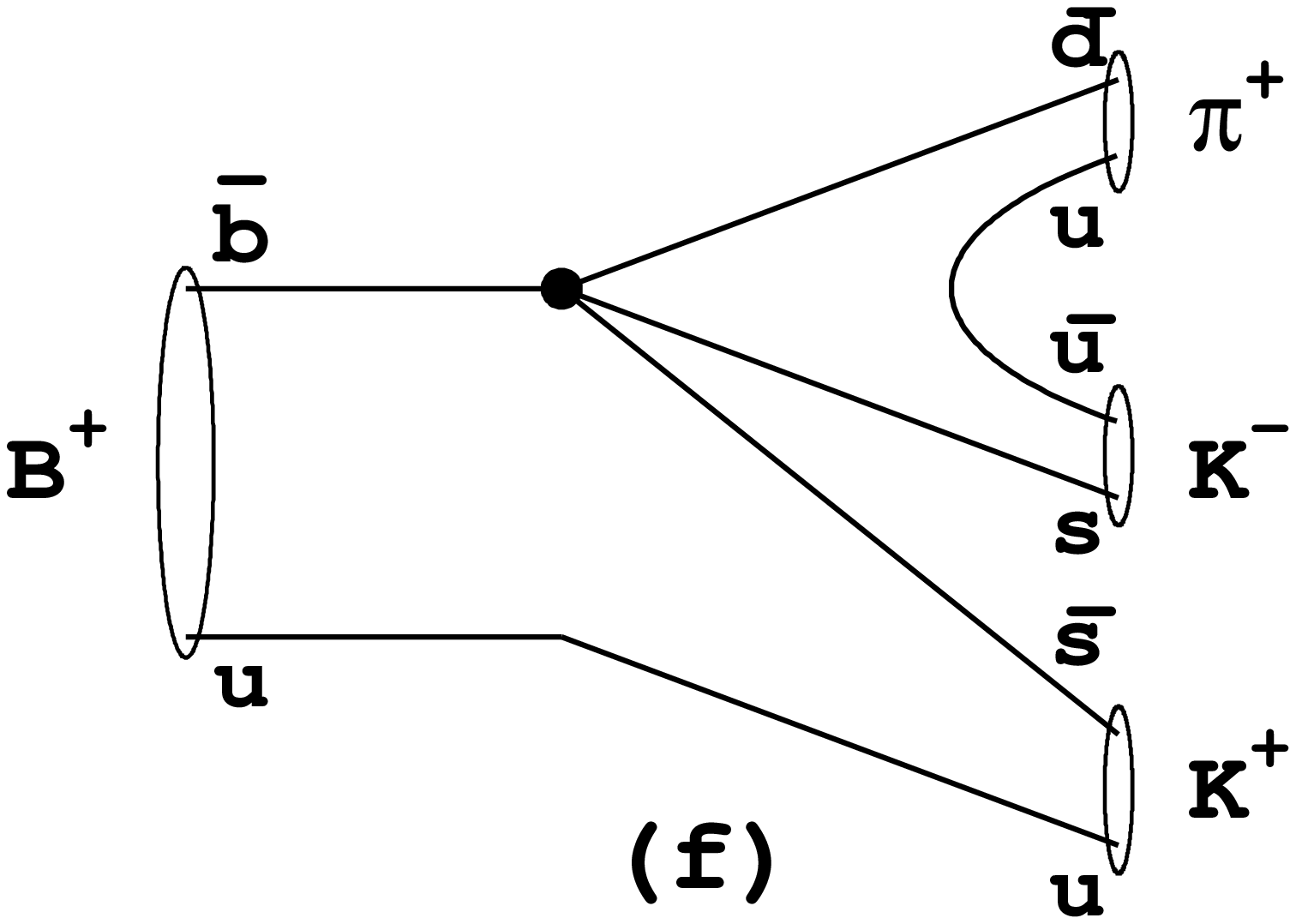}\\
  \end{minipage}
  \begin{minipage}[c]{1.0\textwidth}
  \caption{Diagrams for $B^+\to K^+K^+K^-$ decay:
           (a) $b\to s$ penguin; (b) and (c) $b\to u$ trees, and
           for $B^+\to K^+K^-\pi^+$ decay: (d) and (e) $b\to u$ trees;
           (f) $b\to d$ penguin.}
  \label{fig:diagrams}
  \end{minipage}
\end{figure}

The decays of $B$ mesons to three-body $Khh$ final states can be described by
$b\to u$ tree-level spectator and  $b\to s(d)g$ one-loop penguin diagrams.
Although electroweak penguins, $b\to u$ $W$-exchange, and annihilation diagrams
can also contribute to these final states, they are expected to be much
smaller and we neglect them in the following discussion. $B$ meson decays to
final states with odd numbers of kaons ($s$-quarks) are expected to proceed
dominantly via the $b\to sg$ penguin transition since, for these states, the
$b\to u$ tree contribution  has an additional CKM suppression. In contrast,
$B$ decays to three-body final states with two kaons proceed via the $b\to u$
tree and $b\to dg$ penguin transitions with no $b\to sg$ penguin contribution.
This allows us to make a rough estimate of the $b\to u$ tree contribution to
final states with three kaons via the analysis of $KK\pi$ final states. This
is illustrated for the $\bckkk$ decay in Fig.~\ref{fig:diagrams}, where the
dominant $b\to s$ penguin graph is shown in Fig.~\ref{fig:diagrams}(a). The
$b\to u$ tree graph (Fig.~\ref{fig:diagrams}(b)) has an additional Cabibbo
suppression from the $W^+\to \bar{s}u$ vertex. The corresponding diagram
without Cabibbo suppression ($W^+\to \bar{d}u$) shown in
Fig.~\ref{fig:diagrams}(d) is expected to be the dominant contributor to the
$\kkpos$ final state. Assuming factorization, a quantitative estimate of the
fraction of the $b\to u$ tree amplitude is then provided by the ratio
\begin{equation}
  F \equiv
  \frac{|{\cal{A}}^{KKK}_{b\to u}|^2}{|{\cal{A}}^{KKK}_{\rm total}|^2}\sim
  \frac{{\cal{B}}(B^+\to K^+K^-\pi^+)}{{\cal{B}}(B^+\to K^+K^+K^-)}\times
  \left ( \frac{f_{K}}{f_{\pi}} \right )^2\times\tan^2\theta_C,
\end{equation}
where ${\cal{A}}^{KKK}_{\rm total}$ is the total amplitude for the $\bckkk$
decay and ${\cal{A}}^{KKK}_{b\to u}$ is its $b\to u$ tree contribution. The
$(f_{K}/f_{\pi})^2$ factor, where $f_{\pi} = 131$~MeV and $f_K = 160$~MeV are
the pion and kaon decay constants, respectively, takes into account corrections
for SU(3) breaking effects in the factorization approximation. $\theta_C$ is
the Cabibbo angle ($\sin\theta_C=0.2205\pm0.0018$)~\cite{PDG}. Using the
results for $\bckkpos$ and $\bckkk$ branching fractions from
Table~\ref{tab:defit}, we obtain $F\approx0.022\pm0.005$. Similarly, for $B^0$
decays to $\knkk$ and $\knkp$ final states we find
$F\approx0.023\pm0.013~(<0.037)$,
where the value in brackets is obtained using the upper limit for the $\bnknkp$
branching fraction. Thus, the contribution of the $b\to u$ tree transition is
expected to be at the level of a few percent in branching fraction or at the
level of $10-15$\% in amplitude. In contrast to two-body decays, where the
magnitude of the interference term depends only on the relative phase between
two amplitudes, in three-body decay, for the interference term to be maximal
not only must the relative phase between the two amplitudes be 0 (180) degrees
at any point of the phase space, but these two amplitudes should also have
identical behavior over the phase space. It seems unlikely that all these
conditions will be satisfied since $b\to s$ penguin and $b\to u$ tree
amplitudes are quite different in nature. We conclude that the above value is
a rather conservative estimate of the $b\to u$ contribution.

Let us now consider the CP content of the $\kskk$ state in neutral $B$ decays.
The CP-parity of the $\kskk$ three-body system is $(-1)^{l'}$, where $l'$ is
the orbital angular momentum of the $\kpkm$ pair relative to the remaining
neutral
kaon. Since the total angular momentum of the $\knkk$ system is zero, $l'$ is
equal to the relative orbital angular momentum, $l$, of the two charged kaons.
Thus,
the relative fraction of CP-even and CP-odd states in the $\kskk$ final state
is determined by a fraction of states with even and odd orbital angular momenta
in the
$\kpkm$ system. This fraction could be determined by amplitude analysis of the
$\bnkskk$ Dalitz plot. Such an analysis requires high statistics and cannot be
performed with the available data. Instead, we use isospin relations between
the different three-kaon final states to determine the relative fraction of
CP-even and CP-odd states. Noting that the dominant $b\to sg$ penguin
transition is an isospin conserving process, and neglecting the isospin
violating $b\to u$ tree and $b\to dg$ penguin contributions, we can write the
following relations:
\begin{equation}
  {\cal{B}}(\bnknkk) = {\cal{B}}(\bckknkn)\times
  \frac{\tau_{B^0}}{\tau_{B^+}};
  \label{eq:rel_1}
\end{equation}
\begin{equation}
  {\cal{B}}(\bnknknkn) = {\cal{B}}(\bckkk)\times
  \frac{\tau_{B^0}}{\tau_{B^+}},
  \label{eq:rel_2}
\end{equation}
where the factor $\tau_{B^+}/\tau_{B^0} = 1.091\pm0.023\pm0.014$~\cite{Blife}
takes into account the difference in total widths of charged and neutral $B$
mesons. Being a mirror reflection of each other in isospin space, $\bnknkk$
and $\bckknkn$ decays should have not only equal partial widths
(Equation~\ref{eq:rel_1}), but also the same decay dynamics; specifically, the
fraction of a certain angular momentum state in the $\kpkm$ system in the
$\knkk$ final state should be the same as that for the $\knkn$ system in the
$\knknk$ final state. The $\bckknkn$ decay produces three different observable
states: $\ksksk$, $\klklk$ and $\ksklk$. The relative fractions of these states
depend on the relative fractions of even and odd orbital angular momentum
states in
the $\knkn$ system. Bose statistics requires that the $\knkn$ wave function
be symmetric (and, therefore, CP-even), independently of the relative orbital
angular momentum, $l$, in the system of neutral kaons. As a result, a $\knkn$
system
with $l$-even can only decay to $\ksks$ or $\klkl$ final states (with equal
fractions), while a $K^0\bar{K}^0$ system with $l$-odd can only decay to the
$\kskl$ final state (with an accuracy up to CP violation effects in the system
of neutral kaons).

In this analysis we reconstruct only the $\ksksk$ component of the $\knknk$
final state. Measuring the $\bnknkk$ and $\bcksksk$ branching fractions and
using the isospin relation (\ref{eq:rel_1}) we can determine the
parameter~$\alpha^2$,
\begin{equation}
  \alpha^2 =
  2\frac{{\cal{B}}(\bcksksk)}{{\cal{B}}(\bnknkk)}\times
  \frac{\tau_{B^0}}{\tau_{B^+}} = \frac{N_{\ksksk}}{N_{\kskk}}\times
  2\frac{\varepsilon_{\knkk}}{\varepsilon_{\ksksk}}\times
  \frac{\tau_{B^0}}{\tau_{B^+}}.
  \label{eq:alpha_1}
\end{equation}
Here, $\alpha^2$ characterizes the fraction of states with even orbital angular
momenta
in the $\knkn$ system in the three-body $\knknk$ final state. Due to isospin
symmetry it also gives the fraction of states with even angular momenta in the
$\kpkm$ system for the $\kskk$ final state. With the results given in
Table~\ref{tab:defit}, we obtain: $\alpha^2 = 0.86\pm0.15\pm0.05$, where we use
the expression for $\alpha^2$ in terms of signal yields ($N$) and
reconstruction efficiencies ($\varepsilon$) instead of branching fractions to
reduce the systematic error due to uncertainty in branching fractions of the
calibration modes. Note that the $\kskk$ final state includes the $\phi K^0_S$
state which is CP-odd. We remove $B^0\to\phi K^0_S$ events by requiring
$|M(\kpkm)-M_{\phi}|>15$~MeV/$c^2$; the number of remaining $\kskk$ events is
$123\pm14$. The $\alpha^2$ value for the remaining events is:
$\alpha_{\rm non~\phi}^2=1.04\pm0.19\pm0.06$. The fact that the
$\alpha_{\rm non~\phi}^2$ value is close to unity provides evidence for the
dominance of the CP-even component in the three-body $\kskk$ final state when
the $\phi K^0_S$ intermediate state is excluded.

From this analysis we conclude that the three-body $\bnkskk$ decay can be
useful for the measurement of CP violation in $b\to sg$ penguin dominated
decays. Measurements of the time-dependent CP asymmetry in the $b\to sg$
penguin dominated $B^0\to\eta'K^0_S$, $B^0\to \phi K^0_S$ and $\bnkskk$ decays
are reported in Ref.~\cite{b2sss}.


\section{Conclusion}

  In conclusion, we have measured branching fractions for charmless $B$ meson
decays to the $\kppos$, $\knpp$ and three-kaon $\kkk$, $\knkk$, $\ksksk$ and
$\ksksks$ final states. We also observe $4\sigma$ evidence for the signal in
the $\kkpos$ final state that is expected to be dominated by the $b\to u$ tree
transition. We do not see any signal in $\bckppss$, $\bckkpss$, $\bnknkp$ and
$\bcksksp$ decays and present 90\% CL upper limit for their branching
fractions. All the results on three-body branching fractions are summarized in
Table~\ref{tab:defit}. The results presented in this work are in good agreement
with our previous measurements~\cite{garmash} and with those reported by the
BaBar~\cite{aubert} and CLEO~\cite{eckhart} experiments. This paper does not
update the results for quasi-two-body states obtained using the simplified
technique in our previous analysis~\cite{garmash} because of the large model
error of this technique. The extraction of branching fractions for exclusive
quasi-two-body intermediate states in the observed three-body signals requires
a full amplitude analysis of the corresponding Dalitz plots and is currently
in progress.

An isospin analysis of the charmless $B$ meson decays to three-kaon final
states suggests the dominance of the CP-even component in the three-body
$\kskk$ final state when the $\phi K^0_S$ intermediate state is removed. Using
this final state increases the statistics available for measurements of CP
violation in $b\to s$ penguin dominated decays by a factor of four compared
to $B^0\to\phi K^0_S$.

\section*{Acknowledgments}

We wish to thank the KEKB accelerator group for the excellent
operation of the KEKB accelerator.
We are grateful to V.~Chernyak for fruitful discussions on the theoretical
aspects of the analysis.
We acknowledge support from the Ministry of Education,
Culture, Sports, Science, and Technology of Japan
and the Japan Society for the Promotion of Science;
the Australian Research Council
and the Australian Department of Education, Science and Training;
the National Science Foundation of China under contract No.~10175071;
the Department of Science and Technology of India;
the BK21 program of the Ministry of Education of Korea
and the CHEP SRC program of the Korea Science and Engineering Foundation;
the Polish State Committee for Scientific Research
under contract No.~2P03B 01324;
the Ministry of Science and Technology of the Russian Federation;
the Ministry of Education, Science and Sport of the Republic of Slovenia;
the National Science Council and the Ministry of Education of Taiwan;
and the U.S.\ Department of Energy.


\begin{thebibliography}{99}

\bibitem{b2hhhcp}{N.G.~Deshpande, N.~Sinha and R.~Sinha,
         Phys. Rev. Lett. {\bf 90}, 061802 (2003).}
%
\bibitem{huiti-1}{K.~Huitu, C.D.~Lu, P.~Singer, D.X.~Zhang,
         Phys. Rev. Lett. {\bf 81}, 4313 (1998).}
%
\bibitem{huiti-2}{K.~Huitu, C.D.~Lu, P.~Singer, D.X.~Zhang,
         Phys. Lett.  B {\bf 445}, 394 (1999).}
%
\bibitem{fajfer}{S.~Fajfer and P.~Singer, Phys. Rev. D {\bf 62}, 117702 (2000);
         S.~Fajfer and P.~Singer, Phys. Rev. D {\bf 65}, 017301 (2002).}
%
\bibitem{garmash}{A.~Garmash {\it et al.} (Belle Collaboration),
         Phys. Rev. D {\bf 65}, 092005 (2002).}
%
\bibitem{aubert}{
         B.~Aubert {\it et al.} (BaBar Collaboration), hep-ex/0206004.}
%
\bibitem{eckhart}{
         E. Eckhart {\it et al.} (CLEO Collaboration), hep-ex/0206024.}
%
\bibitem{PDG}{K.Hagiwara et al., Phys. Rev. D {\bf 66}, 010001 (2002) and 2003
         off-year partial update for the 2004 edition (http://pdg.lbl.gov).}
%
\bibitem{KEKB}{S.~Kurokawa, Nucl. Instr. and Meth. A {\bf 499}, 1 (2003).}
%
\bibitem{Belle}{A.~Abashian {\it et al.} (Belle Collaboration),
         Nucl. Instr. and Meth. A {\bf 479}, 117 (2002).}
%
\bibitem{GEANT}{R.Brun {\it et al.},
         GEANT 3.21, CERN Report DD/EE/84-1, 1984.}
%
\bibitem{fisher}{ R.A.~Fisher, Ann. Eugenics {\bf 7}, 179 (1936);
         M.G.~Kendall and A.~Stuart, {\it The Advanced Theory of Statistics},
         2nd ed. (Hafner Publishing, New York, 1968), Vol.III.}
%
\bibitem{VCal}{D.M.~Asner {\it et al.} (CLEO Collaboration),
         Phys. Rev. {\bf D53}, 1039 (1996).}
%
\bibitem{qqcleo}{Events are generated with the CLEO group's QQ program
         (http://www.lns.cornell.edu/public/CLEO/soft/QQ).}
%
\bibitem{psiveto}{For $J/\psi(\psi(2S))$ rejection, we use the muon mass
         hypothesis to calculate the invariant mass of the two tracks.}
%
\bibitem{ArgusF}{H. Albrecht {\it et al.} (ARGUS Collaboration),
         Phys. Lett. B {\bf 229}, 304 (1989).}
%
\bibitem{ul-calc}{For the $\kppss$, $\kkpss$ and $\ksksp$,  modes, we
         treat the fit result as a Gaussian measurement and quote 90\% CL
         upper limits using the procedure of G.J.~Feldman and R.D.~Cousins,
         Phys. Rev. D {\bf 57}, 3873 (1998). For $\kskp$, we quote only
         the upper edge of the Feldman-Cousins interval: this corresponds to
         an upper limit of confidence level greater than 90\%. For the $\kkpos$
         mode, where we observe a $4\sigma$ signal, we quote $\mu+1.28\sigma$,
         where $\mu$ and $\sigma$ are the central value and the error returned
         by the fit.}
%
\bibitem{CP_phi1}{K.~Abe {\it et al.} (Belle Collaboration),
         Phys. Rev. D {\bf 66}, 071102 (2002);
         B. Aubert {\it et al.} (BaBar Collaboration), 
	 Phys. Rev. Lett. {\bf 89}, 201802 (2002).}
%
\bibitem{phik} B.~Aubert {\it et al.} (BaBar Collaboration),
         Phys. Rev. Lett. {\bf 87}, 151801 (2001);
         K.-F.~Chen, A.~Bozek {\it et al.} (Belle Collaboration),
         hep-ex/0307014.
%
\bibitem{kskk-cp}
         K.~Abe {\it et al.} (Belle Collaboration), hep-ex/0208030.
%
\bibitem{GLNQ}
         At the time of the preparation of this manuscript, papers
         by Y.~Grossman, Z.~Ligeti, Y.~Nir, and H.~Quinn (hep-ph/0303171)
	 and by M.~Gronau and J.L.~Rosner (Phys. Lett. B {\bf 564}, 90 (2003))
         appeared that present more general formulations for isospin analyses
         of three kaon final states.
%
\bibitem{Blife}{K.~Abe {\it et al.} (Belle Collaboration),
         Phys. Rev. Lett. {\bf 88}, 171801 (2002).}
%
\bibitem{b2sss}{K.~Abe {\it et al.} (Belle Collaboration),
         Phys. Rev. D {\bf 67}, 031102(R) (2003).}
%
\end{thebibliography}
\end{document}